\documentclass[fleqn,usenatbib,useAMS]{mnras}

\usepackage{newtxtext,newtxmath}
\usepackage{braket}
\usepackage{natbib}
\usepackage{booktabs}
\usepackage{multirow}
\usepackage{ulem}
\usepackage[dvipsnames]{xcolor}

\usepackage[T1]{fontenc}

\DeclareRobustCommand{\VAN}[3]{#2}
\let\VANthebibliography\thebibliography
\def\thebibliography{\DeclareRobustCommand{\VAN}[3]{##3}\VANthebibliography}

\usepackage{graphicx}	
\usepackage{amsmath}	
\usepackage{changes}                                         

\newcommand{\ms}{\mbox {$M_{\odot}$}}

\title[A holistic model of magnetic braking -- I]{Towards a holistic magnetic braking model from the evolution of cataclysmic variables to stellar spin-down -- I: the spin-down of fully convective M-dwarfs}
\author[Sarkar, Yungelson \& Tout]{
Arnab Sarkar$^{1}$\thanks{E-mail: as3158@cam.ac.uk},
Lev Yungelson$^{2}$\thanks{E-mail: lev.yungelson@gmail.com}
and  Christopher A. Tout$^{1}$\thanks{E-mail: cat@ast.cam.ac.uk}
\\
\\
$^{1}$Institute of Astronomy, The Observatories, Madingley Road, Cambridge CB3 OHA, UK
\\
$^{2}$Institute of Astronomy of the Russian Academy of Sciences, 48 Pyatnitskaya Str.,119017 Moscow, Russia
}

\date{Accepted XXX. Received YYY; in original form ZZZ}

\pubyear{2023}

\begin{document}
\label{firstpage}
\pagerange{\pageref{firstpage}--\pageref{lastpage}}
\maketitle

\begin{abstract}
We extend a magnetic braking (MB) model, which has been used earlier to address the evolution of cataclysmic variables, to address the spin period $P_\mathrm{spin}$ evolution of fully convective M dwarf (FCMD) stars. The MB mechanism is an $\alpha-\Omega$ dynamo, which leads to stellar winds that carry away angular momentum. We model our MB torque such that the FCMDs experience a MB torque, approximately scaling as $P_\mathrm{spin}^{-1}$ at shorter periods, before transitioning into a Skumanich-type MB torque, scaling as $P_\mathrm{spin}^{-3}$. We also implement a parametrized reduction in the wind mass loss owing to the entrapment of winds in dead zones. We choose a set of initial conditions and vary the two free parameters in our model to find a good match of our spin trajectories with open clusters containing FCMDs such as NGC2547, Pleiades, NGC2516 and Praesepe. We find that our model can explain the long spin periods of field stars and that a spread in spin distribution persists till over 3 Gyr. \textcolor{black}{An advantage of our model is in relating physically motivated estimations of the magnetic field strength and stellar wind to properties of the stellar dynamo, which other models often remain agnostic about.} We track the spin dependence of the wind mass losses, Alfvén radii and surface magnetic fields and find good agreement with observations. We discuss the implications of our results on the effect of the host FCMD on any orbiting exoplanets and our plans to extend this model to explain solar-like stars in the future. 
\end{abstract}


\section{Introduction} 
Magnetic braking (MB) is a mechanism operating in a variety of stellar systems, in which magnetized winds from the star carry away mass and, as a consequence, angular momentum. This mechanism operates in single stars, which spins them down over time \citep{1967ApJ...148..217W}. Stellar spin-down is heavily dependent on the stellar mass $M_\ast$, such that stars with radiative envelopes ($M_\ast\gtrsim 1.4M_\odot$) do not spin down appreciably over time whereas less massive stars, such as our Sun, with convective envelopes, spin down with time \citep[see, e.g.,][and references therein.]{1992ApJ...390..550M}. \textcolor{black}{ This illustrates the significance of the presence of a convective envelope on magnetic braking, and serves as a motivation to study a possible dynamo mechanism in the convective envelope leading to MB.} 

Stars in binary systems similarly experience magnetic braking so that, owing to tidal locking between the spin of the primary star and the orbit of the system, MB of either star leads to the loss of orbital angular momentum. \textcolor{black}{Depending on the masses of components, its orbital period and the extent of exhaustion of hydrogen in the core of the primary component at the commencement of RLOF, the binary may further evolve to shorter or longer orbital periods 
\citep{1985SvAL...11...52T,1988A&A...191...57P}.} This phenomenon, in conjunction with orbital angular momentum loss owing to gravitational  {waves} radiation \citep{1967AcA....17..287P}, drives the evolution of close binaries such as cataclysmic variable (CV) stars  {and low-mass X-ray binaries}.

Tests of the validity of various MB mechanisms can be made from observations. For CVs, the salient features are (i) a dearth of mass-transferring systems with orbital periods $2\lesssim P_\mathrm{orb}/\mathrm{hr}\lesssim 3$, known as the period gap
   \citep{2003cvs..book.....W}, (ii) a spike in the number of observed systems with $P_\mathrm{orb}\approx 80\, \mathrm{min}$, known as the period minimum spike
\citep{1999MNRAS.309.1034K, Gnsicke2009} and (iii)  {a discrepancy of unclear origin} between the observed average white dwarf 
accretors mass of about $0.8 \,M_\odot$ and 
a significantly lower one for isolated white dwarfs  of about $0.6M_\odot$ \citep[e.g.,][]{Wijnen2015,Pala2021}. Out of these features, attempts have been made to explain the first two with various MB paradigms \citep[e.g.,][]{1981A&A...100L...7V,1989ApJ...345..972T,Knigge2006, 2009MNRAS.397.2208Z,Knigge2011} \textcolor{black}{as well as by a decline of the magnetic field strength due to its increasing complexity (\citealt{1989ApJ...345..972T} and the model of \citealt{2018ApJ...868...60G} for illustration of this suggestion).}


Recently, \citet{2022MNRAS.513.4169S}, building on the model of \cite{Zangrilli1997}, explained the existence
of the period gap and the period minimum using a MB mechanism with two $\alpha-\Omega$ dynamos operating in the donor ($M_\ast\lesssim 1.4M_\odot$), one at the boundary 
of the radiative core and the convective envelope and the other in the convective envelope. The dynamo at the boundary layer ceases to operate once mass loss from the donor reduces $M_\ast$ below $\approx 0.3\,\ms$ at $P_\mathrm{orb}\approx 3\, \mathrm{hr}$, thus causing 
the contraction of the donor and  cessation of mass transfer. However, angular momentum is still lost from the system owing to gravitational radiation and MB by the convective dynamo. Mass transfer 
resumes at $P_\mathrm{orb}\approx 2\, \mathrm{hr}$ at the bottom of the period gap.
 As the fully convective star undergoes further mass loss, there is an enhancement in the convective dynamo which leads to a period minimum at $P_\mathrm{orb}\approx 80 \,\mathrm{min}$, as the donor becomes degenerate and expands as it loses mass. If angular momentum were lost only by GR then this period minimum would be at $P_\mathrm{orb}\approx 65\, \mathrm{min}$ \citep{1993A&A...271..149K}. It is raised because the higher mass-loss rate forces the donor to respond adiabatically and expand.
This MB paradigm for CVs, 
implemented  for the treatment of helium-star and evolved-donor AM CVn stars 
evolution \citep{2023MNRAS.519.2567S,Sarkar2023}, has given results that are in general agreement with observations. 

The spins of single-star systems show many salient features that may be explained with an apt MB formalism. Observations of open clusters (OCs) of known ages reveal that the member stars show a bimodality in their rotation rates, such that the same OC contains a population of fast as well as slow rotators \citep{2003ApJ...586..464B}. The bimodality is independent of the mass of the star and has been observed in both solar-like stars \citep{2009ApJ...695..679M, 2011ApJ...733L...9M} which possess convective envelopes and radiative cores, as well as fully convective M dwarfs \citep{Douglas2016, Newton2016}. However, the time for which this bimodality persists is mass-dependent, with older clusters and field stars (with ages greater than $1\,\mathrm{Gyr}$) still showing such a bimodality in their late-M dwarf subset when earlier-type stars have spun down significantly to become slow rotators \citep{Pass2022,Popinchalk2021}. A MB model to explain some of these features was first posited by \cite{Brown2014} with a model in which there is a sudden, stochastic and mass-dependent coupling of the stellar dynamo with the wind that causes enhanced MB. This model was used to explain the bimodality of rotation rates in OCs. Owing to the random, and potentially nonphysical nature of this model, \cite{Garraffo2018} improved upon it by arguing, based on simulations by \cite{Garraffo2015,Garraffo2016}, that the extent of angular momentum loss owing to MB is dependent on the morphology of the magnetic field, in such a way that fast rotators have more of their magnetic flux in higher order magnetic multipoles. This leads to a reduction in the MB efficiency, causing fast rotators to lose less angular momentum than slow rotators. \textcolor{black}{However, \cite{2019ApJ...886..120S} showed that in most cases multi-polar effects do not affect MB efficiency, with increased possible contribution at larger Rossby numbers \citep{2022ApJ...933L..17M, 2023ApJ...948L...6M}.}

In this paper we extend our MB mechanism for CVs, the Double Dynamo (DD) model, to address the problem of stellar spin down, focusing only on fully convective M dwarfs or FCMDs, with $M_\ast\lesssim0.35M_\odot$. For CVs, the donor star becomes fully convective below the period gap, when the $P_\mathrm{orb}\lesssim2\,\mathrm{hr}$. \cite{2022MNRAS.513.4169S} posited that, at this stage, only the convective dynamo causes MB in the donor star. This convective dynamo was first introduced in \citet[hereinafter TP]{1992MNRAS.256..269T} in the context of contracting pre-main-sequence (PMS) stars, and extended by \cite{1995MNRAS.273..146R} to explain common envelope evolution in the formation of CVs. 
Because we are focusing only on FCMDs, we extend this dynamo implementation of TP. The general motivation behind the use of the DD model in the field of CV evolution was to explain the orbital properties of the system and not the features of the donor star such as its wind mass loss and magnetic field, primarily because we cannot usually infer them from observations of non-magnetic CVs. 
However, for single-star evolution, modelling these properties in detail is essential because M-dwarf winds, magnetic fields, Alfvén radii, etc. can be inferred from observations \citep{2018MNRAS.475L..25V}. So we construct our MB model here from the general ideas of TP but derive all the equations from scratch. We also refrain from referring to this MB model as the DD model because of significant changes implemented in its construction since the work of \cite{2022MNRAS.513.4169S}.

This paper is structured as follows. In Section~\ref{sec:math} we explain the physics of our model in detail. In Section~\ref{sec:spind} we show how our model works for FCMDs and its dependence on the initial conditions. In Section~\ref{sec:results} we show how our model explains the important observational trends in the spin-down of FCMDs. We discuss the implications of our results in Section~\ref{sec:discussion} and conclude in Section~\ref{sec:conclusions}.

\section{The magnetic braking model}
\label{sec:math}
We base our dynamo model on the work of TP who presumed that rotation and convection in a PMS star lead to a dynamo action and create magnetic fields. Magnetic flux is lost from the star owing to magnetic buoyancy, which deposits energy at the stellar surface and this energy escapes the star in the form of a magnetic stellar winds, which carry away angular momentum. 

\subsection{The dynamo mechanism}
\label{subs:dynamo}
It has been known for a while that the base of the solar convective zone has a higher angular velocity than the surface, while the middle region rotates with a lower angular velocity than the surface \citep{Duvall1984}. 
 In addition to this, there is a latitudinal variation in the angular velocity of the convective zone \citep{1985ApJ...297..787D}. In other words, there is differential rotation in the solar convective zone. We assume that a similar angular velocity variation exists in FCMDs. For our simple one-dimensional dynamo, we assume that the star rotates as a solid body and parametrize the differential rotation with an average across the star $\Delta \Omega$. We further assume that 
$\Delta \Omega \approx \Omega$ so that the differential rotation reduces when the star spins down. This is the $\Omega$ term in our $\alpha-\Omega$ dynamo, which converts poloidal fields into toriodal fields. The dynamo is completed when the regeneration term or the $\alpha$ term in the $\alpha-\Omega$ dynamo converts toroidal fields into poloidal fields. We assume that our dynamo reaches a steady state fairly quickly such that magnetic fields are created and destroyed at the same rate. So we use the equilibrium equations (4.1) and (4.2) of TP,
\begin{center}
\begin{equation}
\label{dyn1}
\frac{\mathrm{d}B_\mathrm{\phi}}{\mathrm{d}t} =\Delta\Omega B_\mathrm{p} - \frac{B_\mathrm{\phi}}{\tau_\mathrm{\phi}} = 0
\end{equation}
\end{center}
and
\begin{center}
\begin{equation}
\label{eq:dyn2}
\frac{\mathrm{d}A_\mathrm{\phi}}{\mathrm{d}t} =\Gamma B_\mathrm{\phi} - \frac{A_\mathrm{\phi}}{\tau_\mathrm{p}} = 0,
 \end{equation}
\end{center}
where $B_\mathrm{\phi}$ is an average toroidal component of the magnetic field in the star, $A_\mathrm{\phi}$ is an average azimuthal component of the magnetic vector potential, such that the poloidal component of the magnetic field $B_\mathrm{p}\approx A_\mathrm{\phi}/R_\ast$,  $\tau_\mathrm{\phi}$ and $\tau_\mathrm{p}$ are the time-scales on which the toroidal and poloidal magnetic field components are destroyed, 
and $\Gamma$ is the dynamo regeneration term.
 Equation (\ref{eq:dyn2}) can then be written as
\begin{center}
\begin{equation}
\label{eq:dyn3}
\frac{\mathrm{d}B_\mathrm{p}}{\mathrm{d}t} =\frac{\Gamma}{R_\ast} B_\mathrm{\mathrm{\phi}} - \frac{B_\mathrm{p}}{\tau_\mathrm{p}} = 0.
 \end{equation}
\end{center}
We also note that we expect the shear term to act more rapidly than the regeneration term and therefore expect $B_\mathrm{\phi}\gg B_\mathrm{p}$ and write $B_\mathrm{p}\approx \epsilon B_\mathrm{\phi}$ where $\epsilon\ll 1$. The $\alpha$ effect was discussed in the hydromagnetic dynamo model of \cite{1955ApJ...122..293P} who suggested that convective eddies, moving through toroidal fields experience cyclonic turbulence, creating loops of poloidal field from the toroidal field. These loops merge to form large-scale poloidal fields from toroidal fields. An expression for $\Gamma$ is \citep{1955ApJ...122..293P, 1981ARA&A..19..115C}
\begin{center}
\begin{equation}
\label{eq:Gamma}
\Gamma \approx \tau_\mathrm{t}v_\mathrm{t}\omega_\mathrm{t} = \gamma v_\mathrm{c},
 \end{equation}
\end{center}
where $\tau_\mathrm{t}$ is the turbulent turnover time-scale, $v_\mathrm{t} \approx v_\mathrm{c}$ is the velocity of turbulent cells, $v_\mathrm{c}$ is the convective velocity and $\omega_\mathrm{t}\propto\Omega$ is the vorticity of the eddies parallel to $v_\mathrm{c}$. The angle 
 {through} 
which each eddy rotates for each 
 {turn of the star} is given by $\omega_\mathrm{t}\tau_\mathrm{t}$. TP assumed this to be of the order $2\pi$ for all $\Omega$s. As a result, their $\Gamma$ was a constant for a given $v_\mathrm{c}$. However, it has been observed that $B_\mathrm{p}\propto\Gamma$ has a dependence on $\Omega$ for M dwarf stars \citep{2009ApJ...692..538R}. In addition, a $\Gamma$ that is independent of rotation fails to explain the plethora of observed magnetic fields in M dwarfs \citep{2021A&ARv..29....1K}. Owing to these arguments, and the fact that $\omega_\mathrm{t}\propto\Omega$, we write
\begin{center}
\begin{equation}
\label{eq:gamma}
\gamma = f_1 \frac{M_\ast}{0.35 M_\odot} \sqrt{\frac{R_\ast}{g}} \Omega,
 \end{equation}
\end{center}
where $f_1$ is a free parameter of the order unity with which we can calibrate our model. Here ${M_\ast}/{0.35 M_\odot}$ is the mass that undergoes this instability, normalized to $0.35 M_\odot$ which we take to be the maximum mass of a FCMD (however, see Section~\ref{subs:solarlike}). The reasoning behind this is that a higher-mass FCMD has more material that experiences cyclonic turbulence, and as a result, gives rise to a stronger poloidal field (also see Section~\ref{subs:obs}). We assume that turbulent instabilities turn over in the shortest time-scale possible. We take this to be the sound-crossing time over the stellar radius \citep{2002A&A...381..923S}, such that $\tau_\mathrm{t}\approx\sqrt{{R_\ast}/{g}}$, where $g$ is the surface gravity.

\subsection{Wind mass loss calculation}
We follow the arguments of TP and assume that the energy input by shear in the star is comparable to its rotational kinetic energy, such that the rate at which energy is fed into the shear is given by
\begin{center}
\begin{equation}
\label{eq:Lw}
L_\mathrm{w} \approx\frac{ \frac{1}{2} k^2M_\ast R^2_\ast \Omega^2} {\tau_\mathrm{\nu}},
 \end{equation}
\end{center} 
where $k^2\approx0.1$ is the square of the dimensionless radius of gyration and $\tau_\nu$ is the viscous-time scale given by
\begin{center}
\begin{equation}
\label{eq:taunu}
\tau_\mathrm{\nu} = \frac{R_\ast^2}{\nu},
 \end{equation}
\end{center} 
where $\nu$ is the convective viscosity. TP assumed that about 10 percent of the deposited energy escapes the star as winds, and the rest is radiated away\footnote{We point out that a different choice of fractional energy deposition will only modify the calibration of $f_1$ and $f_2$ without affecting our final results in Section~\ref{sec:results}.}. We similarly write
\begin{center}
\begin{equation}
\label{eq:lw2mdot}
0.1L_\mathrm{w} \approx \frac{GM_\ast\Dot{M}_w}{R_\ast},
 \end{equation}
\end{center} 
where $G$ is Newton's gravitational constant. 
We then obtain an expression for the wind mass-loss rate
\begin{center}
\begin{equation}
\label{eq:mlconv0}
\Dot{M}_\mathrm{w} = f_\mathrm{DZ} \frac{1}{200} \frac{R_\ast}{G} \Omega^2 \nu.
\end{equation}
\end{center}
Here $f_\mathrm{DZ}$ is a dimensionless quantity less than 1 which measures the effect of the dead zone on the star's mass-loss rate (Section~\ref{subs:dz}). For a non-rotating star, the form of $\nu$ is given by equation (3.3) of TP
\begin{center}
\begin{equation}
\label{eq:nu0}
\nu \approx \frac{1}{3}v_\mathrm{c}l_\mathrm{c},
\end{equation}
\end{center}
where 
\begin{center}
\begin{equation}
\label{eq:vc}
v_\mathrm{c} \approx \left(\frac{L_\ast R_\ast}{\eta M_\ast}\right)^{1/3}.
 \end{equation}
\end{center}
Here $L_\ast$ is the luminosity of the star, $\eta \approx 3R_\ast/l_\mathrm{c} \approx 30$ is a constant \citep{Campbell1983} and $l_\mathrm{c}$ is the mixing length. For a rapidly rotating star such that its spin period $P_\mathrm{spin}~\lesssim~\tau_\mathrm{c}$, where $\tau_\mathrm{c}=l_\mathrm{c}/v_\mathrm{c}$ is the convective turnover time-scale, only the eddies that turn over in a time shorter than the convective turnover time-scale can contribute to the transport of angular momentum \citep{1977ApJ...211..934G}. So for a rapidly rotating star such that $\Omega\tau_\mathrm{c}\gtrsim 2\pi f_2$, we assume that $\nu$ is reduced by a factor $(2\pi f_2/{\tau_\mathrm{c}\Omega})^p$, where $f_2\gtrsim 1$ is a free parameter of the order unity and $p\geq0$ as given by TP which measures how strongly the convective viscosity is curtailed when $\Omega\tau_\mathrm{c} > 2\pi f_2$ (Eq.~\ref{eq:nu}).
The choice of $p$ is presented in Section~\ref{sec:spind}. With these changes, the expression for $\nu$ is
\begin{center}
\begin{equation}
\label{eq:nu}
\nu \approx \frac{1}{3} v_\mathrm{c}l_\mathrm{c} \:\mathrm{min}\left(\left(\frac{2\pi f_2}{\tau_\mathrm{c}\Omega}\right)^{p},\:1\right),
\end{equation}
\end{center}
and $\Dot{M}_\mathrm{w}$ becomes
\begin{center}
\begin{equation}
\label{eq:mlconv}
\Dot{M}_\mathrm{w} = f_\mathrm{DZ} \frac{1}{600} \frac{R_\ast}{G} \Omega^2 v_\mathrm{c}l_\mathrm{c} \:\mathrm{min}\left(\left(\frac{2\pi f_2}{\tau_\mathrm{c}\Omega}\right)^{p},\:1\right).
\end{equation}
\end{center}

\subsection{Angular momentum loss rate}
Owing to the strong energy deposition by shear, winds at the escape velocity of the star emerge from the stellar surface. These winds are forced to corotate with the star to the Alfvén radius, where  the kinetic energy density in the wind equals the magnetic energy density and so
\begin{center}
\begin{equation}
\label{eq:vA}
v^2_\mathrm{A} \approx \frac{B_\mathrm{p}(R_\mathrm{A})^2}{4\pi\rho_\mathrm{A}},
\end{equation}
\end{center}
where the subscript A denotes the Alfvén surface. Assuming a spherically symmetric wind, we obtain
\begin{center}
\begin{equation}
\label{eq:rhoA}
\rho_\mathrm{A} = \frac{\Dot{M}_\mathrm{w}}{4\pi R_\mathrm{A}^2 v_\mathrm{A}}.
\end{equation}
\end{center}
We now assume that the magnetic field falls off as a dipole so that 
\begin{center}
\begin{equation}
\label{eq:bpR}
B_\mathrm{p}(R) = B_\mathrm{p}(R_\ast)\left(\frac{R_\ast}{R}\right)^3,
\end{equation}
\end{center}
and that the velocity at the Alfvén surface is of the order of the escape velocity of the star (see also \citealt{2012ApJ...746...43R}) 
\begin{center}
\begin{equation}
\label{eq:vA2}
v_\mathrm{A}\approx v_\mathrm{esc} = \sqrt{\frac{2GM_\ast}{R_\ast}}.
\end{equation}
\end{center}
With these assumptions we get
\begin{center}
\begin{equation}
\label{eq:ra}
R_\mathrm{A} = R_\ast\left(\frac{B_\mathrm{p}(R_\ast)^2R_\ast^2}{\Dot{M}_\mathrm{w}v_\mathrm{esc}}\right)^{1/4}.
\end{equation}
\end{center}
We have ignored the fact that winds are being accelerated by centrifugal force and decelerated by gravity. They are also forced to traverse along the magnetic field lines till the Alfv{\'e}n radius. This will lead to changes in the wind mass loss, which we scale
with $f_\mathrm{DZ}$ in Section~\ref{subs:dz}. The dynamo mechanism leads to a surface poloidal field which is given by equation (4.10) of TP:
\begin{center}
\begin{equation}
\label{eq:bpconv}
B_\mathrm{p}(R_\ast) = 10\gamma v_\mathrm{c}\sqrt{4\pi \rho_\ast},
\end{equation}
\end{center}
where $\rho_\ast$ is the mean
density of the star. Beyond the Alfvén radius, the winds escape freely, carrying away angular momentum. So we write the angular momentum loss by MB as
\begin{center}
\begin{equation}
\label{eq:jdot}
\Dot{J} = - \Dot{M}_\mathrm{w} R_\mathrm{A}^2 \Omega.
\end{equation}
\end{center}

\subsection{Dead zone implementation}
\label{subs:dz}
The combined effects of gravity, magnetism and centrifugal force give rise to dead zones, posited by \citet[hereinafter MS]{1987MNRAS.226...57M}. Their effect is essentially to reduce the stellar magnetic winds that cause braking in the star. For details on the evaluation of the size of dead zones, we refer the reader to Sec.~2 of MS. 
Here we revisit the equations and concepts that we incorporate in our model. Dead zones are regions around a spinning magnetized star where, owing to a stronger effect of magnetic pressure, gas is trapped 
and hence, does not contribute to the wind mass loss, which is consequently curtailed. The surface on which the effects balance out can be found using relation
\begin{center}
\begin{equation}
\label{eq:pbdz}
\frac{B_\mathrm{p}^2}{8\pi} = \rho_\mathrm{d}c_\mathrm{s}^2,
\end{equation}
\end{center}
where $\rho_\mathrm{d}$ is the density at the base of the dead zone, and $c_\mathrm{s}$ is the sound speed of the fluid and is taken to be isothermal for simplicity. With the expression from Eq.~(7) of MS, we can write the function $\rho_\mathrm{d}(r)$ along the equator as
\begin{center}
\begin{equation}
\label{eq:rho_d}
\rho_\mathrm{d}(r) = \rho_\mathrm{d,0}\:\mathrm{exp}\left(-\frac{GM_\ast}{R_\ast c_\mathrm{s}^2}\left(1 - \frac{R_\ast}{r}\right) + \frac{1}{2}\frac{\Omega^2R_\ast^2}{c_\mathrm{s}^2}\left(\frac{r^2}{R_\ast^2} - \frac{R_\ast}{r}\right)\right),
\end{equation}
\end{center}
where $\rho_\mathrm{d,0}$ is the coronal base density. We define $r_\mathrm{DZ}$ as the distance from the centre of the star at which equation (\ref{eq:pbdz}) is satisfied. For a given $r_\mathrm{DZ}$, we define the fraction of open field lines that contribute to the wind mass loss as $f_\mathrm{DZ}= R_\ast/r_\mathrm{DZ}$ as defined by MS and \citet[Eqs.~(32), (33)]{1994MNRAS.268...61L}. From equation (\ref{eq:pbdz}), we obtain an equation for $r_\mathrm{DZ}$,
\begin{center}
\begin{equation}
\label{eq:F}
\textbf{F}\equiv\left(\frac{R_\ast}{r_\mathrm{DZ}}\right)^6 -\frac{1}{\zeta_\mathrm{d}}\,\mathrm{exp}\left(-l_\mathrm{d}\left(1 - \frac{R_\ast}{r_\mathrm{DZ}}\right) + \frac{1}{2}\chi l_\mathrm{d}\left(\frac{r_\mathrm{DZ}^2}{R_\ast^2} - \frac{R_\ast}{r_\mathrm{DZ}}\right)\right) = 0,
\end{equation}
\end{center}
where $\zeta_\mathrm{d} = B_\mathrm{p}(R_\ast)^2/8\pi\rho_\mathrm{d,0}c_\mathrm{s}^2$, $l_\mathrm{d} = GM_\ast/R_\ast c_\mathrm{s}^2$ and $\chi l_\mathrm{d} = \Omega^2 R_\ast^2/c_\mathrm{s}^2$ according to Eq.~(9) of MS. Although our model estimates the surface $B_\mathrm{p}$ at each stellar evolution time step (Eq.~(\ref{eq:bpR})), the biggest uncertainty in this model is the estimation of 
$\zeta_\mathrm{d}$, which depends on $\rho_\mathrm{d,0}$ and $c_\mathrm{s}$, which in turn depend on the temperature of the stellar corona. Owing to uncertainties in estimating these parameters for our M dwarfs, we adopt a simple relation of the form
\begin{center}
\begin{equation}
\label{eq:zeta}
\zeta_\mathrm{d} = 60 \left(\frac{B_\mathrm{p}(R_\ast)}{B_\odot}\right)^2,
\end{equation}
\end{center}
where 60 has been taken from MS and $B_\odot=1\;\mathrm{G}$. With these expressions, we can calculate $f_\mathrm{DZ}$ for a given star, if we know its mass, radius, luminosity and spin.

\section{Stellar spin down with our model}
\label{sec:spind}
With this we can model the spin evolution of FCMDs if we know the time evolution of three parameters, namely $L_\ast$, $R_\ast$, and $M_\ast$. We use the Cambridge stellar evolution code \textsc{STARS} \citep{1973MNRAS.163..279E, 1995MNRAS.274..964P} to obtain these parameters as a function of time. We assume that the winds do not reduce the stellar mass significantly and so keep $M_\ast$ constant with time. Our models with and without the wind's influence on stellar mass yield identical spin evolution results. For a given initial $P_\mathrm{spin,t=0}$, we obtain the spin evolution of the star by Euler integration
\begin{center}
\begin{equation}
\label{eq:omegat}
\Omega{(t+\Delta t)} = \Omega{(t)} + \Dot{\Omega}(t)\Delta t,
\end{equation}
\end{center}
where
\begin{center}
\begin{equation}
\label{eq:omegadot}
\Dot{\Omega}(t) = \frac{-\Dot{J}(t) -\Dot{I}(t)\Omega(t)}{I(t)},
\end{equation}
\end{center}
and $I = k^2M_\ast R^2_\ast$ is the moment of inertia of the star and $\Dot{J}$ is given by Eq.~(\ref{eq:jdot}). We solve for $f_\mathrm{DZ}$ at each time step using a Newton-Raphson method. Now our model of stellar spin-down is ready to be used, if we know $p,\,f_1,$ and $f_2$. In the next section we obtain $p$.

\subsection{Saturated magnetic braking and dead zone implementation}
With Eqs. (\ref{eq:mlconv}) to (\ref{eq:jdot}) we can write 

\begin{figure*}
\centering
\includegraphics[width=\textwidth]{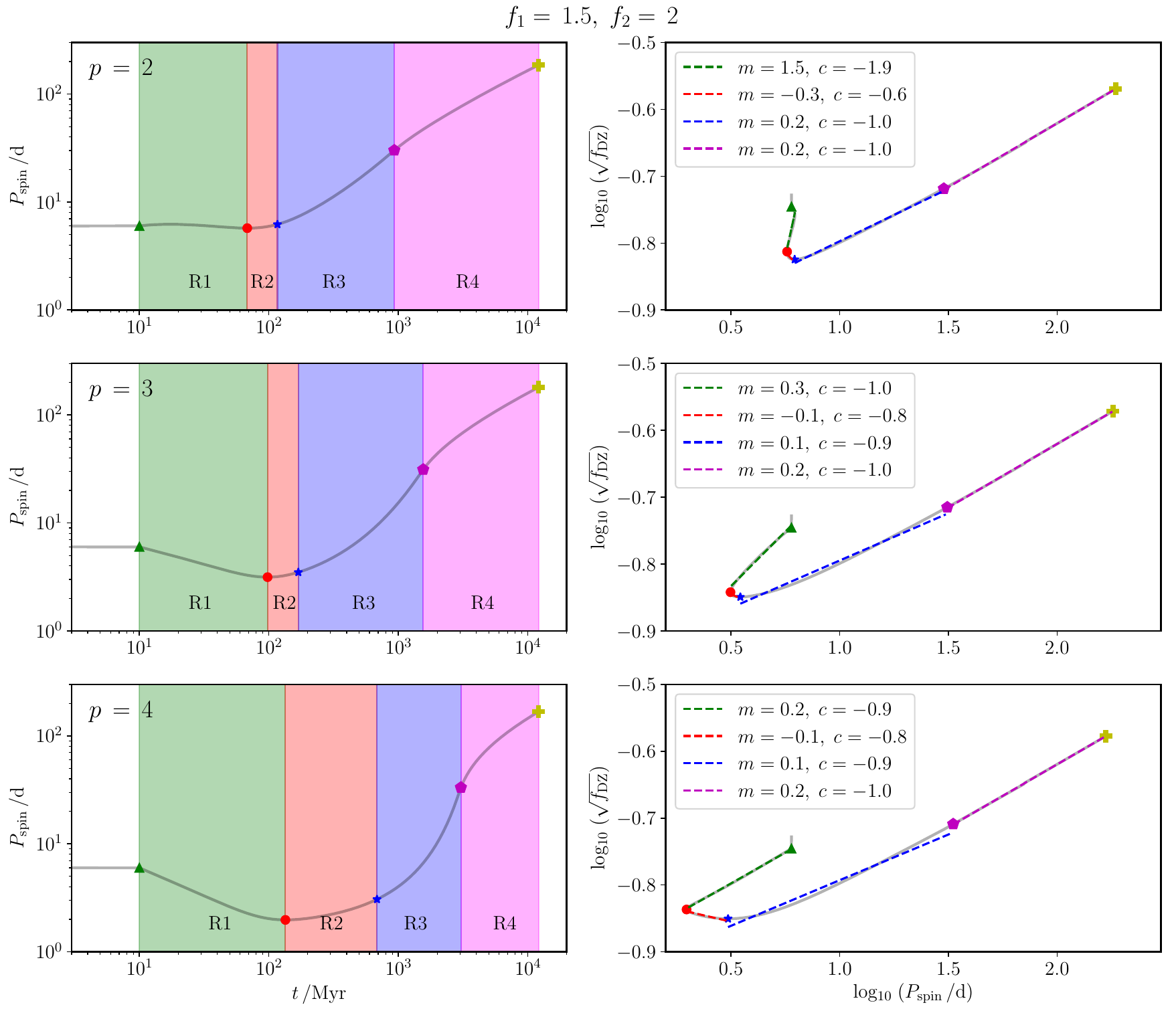}
\caption{The behaviour of our model at various $p$, for a $0.3\,M_\odot$ PMS star with $L_\ast = 1.21L_\odot$, $R_\ast = 1.88R_\odot$, $P_\mathrm{spin,0} = 6\,\mathrm{d}$, $f_1=1.5$, $f_2=\,2$ and $\tau_\mathrm{dl} = 10\,\mathrm{Myr}$. \textit{Left:} The spin evolution of the star with time. The disk-locking time is marked with a green triangle, the time when the star attains its minimum spin period is marked as a red circle. The transition in the behaviour of the dead zone is denoted by the blue star, the end of the evolution by the yellow plus, and the transition of the star from saturated to unsaturated regime by the magenta pentagon. \textit{Right:} The dependence of $f_\mathrm{DZ}$ on $P_\mathrm{spin}$ throughout the spin evolution of the star. The individual fits of each separate regime with a power law of the form $\mathrm{log_{10}}\,\sqrt{f_\mathrm{DZ}} = m\: \mathrm{log_{10}} P_\mathrm{spin} + c$ is shown as dashed lines. \textcolor{black}{The regimes and their respective fits for $f_\mathrm{DZ}$ have been coloured accordingly.}} 
\label{fig:p&fdz}
\end{figure*}

\begin{center}
\begin{equation}
\label{eq:jdotprop}   
\Dot{J} \propto \sqrt{\Dot{M}_\mathrm{w}} \Omega^2
\end{equation}
\end{center}
which, for a given choice of $f_1$ and $f_2$, reduces to
\begin{center}
\begin{equation}
\label{eq:jdotprop_cases}
\Dot{J} \propto
    \begin{cases}
        \sqrt{f_\mathrm{DZ}}\Omega^{3-\frac{p}{2}}, & \text{when   }\; 2\pi f_2/\tau_\mathrm{c}\Omega \; \leq \; 1\; \mathrm{(saturated)},\\
        \\
        \sqrt{f_\mathrm{DZ}}\Omega^{3}, & \text{otherwise}\; \mathrm{(unsaturated)}.
    \end{cases}
\end{equation}
\end{center}
\textcolor{black}{The effect of saturation on the rotational velocities of stars and its association with the magnetic field has been known for a long time (see, e.g., 
\citealt{1988ApJ...333..236K}). \citet{2011ApJ...743...48W}, based on  the sample of about 800
solar and late-type stars with X-ray luminosities and rotation periods, suggested that magnetic braking saturates at smaller spin periods, such that a shallower spin-down torque than a Skumanich-type torque \citep{1972ApJ...171..565S} best reproduces observations. Saturation has been incorporated, for instance, in the magnetic braking models of \cite{Sills2000,Matt2015} and it has been attested in the recent study of \cite{2022MNRAS.517.4916E} wherein it was shown that a saturated braking torque matches best with the observed orbital periods of low-mass detached eclipsing binaries. For this reason, in Eq. (\ref{eq:jdotprop_cases}) we shall call the former scenario the saturated regime and the latter the unsaturated regime because $p>0$ gives a shallower dependence of the torque on the spin. }
Let us assess the behaviour of our model to changes in $p$ and $f_\mathrm{DZ}$. We evolve a $0.3\,M_\odot$ PMS star with $L_\ast = 1.21L_\odot$, $R_\ast = 1.88R_\odot$, such that the star crosses the birth line of \cite{1983ApJ...274..822S} before 3 Myr and an initial period of $6\,\mathrm{d}$ such that the star spins up during its contraction phase and set $f_1=1.5$, $f_2=\,2$ and the disk-locking time $\tau_\mathrm{dl} = 10\,\mathrm{Myr}$. We assume that the star does not go through any spin evolution while $t<\tau_\mathrm{dl}$ \citep{Rebull2004}. We then examine the behaviour of the star till $12\,\mathrm{Gyr}$. This is shown in Fig.~\ref{fig:p&fdz} for $p\in\,\{2,3,4\}$. We mark the end of the disk-locking time with a green triangle. After this, the star spins up\footnote{For $p=2$ this spin-up phase is not visible, and initially there is even a slight spin-down. This is due to a very strong MB spin-down torque (Eq~(\ref{eq:jdotpropp2})) which dominates over the contraction-driven spin-up.} 
until it reaches its minimum spin period, marked with a red circle. \textcolor{black}{ We call this the first regime. It then begins the process of spin down but the dead zone initially grows with an increasing spin period until the blue star. This is the second regime. Subsequently, the dead zone shrinks with increasing period when the system transitions from the saturated to the unsaturated regime owing to equation~(\ref{eq:jdotprop_cases}), denoted by the magenta pentagon. We call this the third regime. From here till the end of its evolution (in the fourth regime), the system spins down under the unsaturated MB torque and the dead zone continues to shrink. The end of the evolution is shown by the yellow plus.} Owing to the different behaviour of $f_\mathrm{DZ}$ during the spin up and the spin down processes, we individually fit each separate regime with a power law of the form $\mathrm{log_{10}}\,\sqrt{f_\mathrm{DZ}} = m\: \mathrm{log_{10}} P_\mathrm{spin} + c$. The dashed green line is our fit in the first regime, the red our fit in the second regime, the blue our fit in the third regime, and the magenta our fit in the fourth regime.  
We can see that torque dependence can be written for the four regimes as
\begin{center}
\begin{equation}
\label{eq:jdotprop2}
\Dot{J} \propto
    \begin{cases}
        \Omega^{3-\frac{p}{2}+m}, & \text{in regimes 1, 2, and 3,}\\
        \\
        \Omega^{3+m}, & \text{in regime 4.}\\
    \end{cases}
\end{equation}
\end{center}
With the best fits from Fig.~\ref{fig:p&fdz}, our three cases approximately become
\begin{center}
\begin{equation}
\label{eq:jdotpropp2}
\Dot{J} \propto
    \begin{cases}
        \Omega^{3.5} & \text{in regime 1, } \\
        \\
        \Omega^{1.7} & \text{in regime 2, }\\
        \\
        \Omega^{2.2} & \text{in regime 3, }\\
        \\
        \Omega^{3.2} & \text{in regime 4, }\\        
    \end{cases}
\end{equation}
\end{center}
for $p=2$,
\begin{center}
\begin{equation}
\label{eq:jdotpropp3}
\Dot{J} \propto
    \begin{cases}
        \Omega^{1.8} & \text{in regime 1, } \\
        \\
        \Omega^{1.4} & \text{in regime 2, }\\
        \\
        \Omega^{1.6} & \text{in regime 3, }\\
        \\
        \Omega^{3.2} & \text{in regime 4, }\\  
    \end{cases}
\end{equation}
\end{center} 
for $p=3$ and
\begin{center}
\begin{equation}
\label{eq:jdotpropp4}
\Dot{J} \propto
    \begin{cases}
        \Omega^{1.2} & \text{in regime 1, } \\
        \\
        \Omega^{0.9} & \text{in regime 2, }\\
        \\
        \Omega^{1.1} & \text{in regime 3, }\\
        \\
        \Omega^{3.2} & \text{in regime 4, }\\  
    \end{cases}
\end{equation}
\end{center} 
for $p=4$. 

 \textcolor{black}{It can be seen that a  Skumanich-like magnetic braking law  with $\Dot{J}\propto\Omega^3$ is not valid throughout the entire spin evolution of the star. The lifetime of a system in regime 4 where it behaves according to the Skumanich law is shorter for smaller $p$. Hence, FCMDs in open clusters  and field FCMDs younger than few Gyr are not following a Skumanich-type MB law, and have a shallower dependence of MB on $\Omega$.} We {notice} that a shallower scaling is seen for lower spins ($P_\mathrm{spin}\lesssim20\,\mathrm{d}$), and as the star spins down further $\Dot{J}$ follows a Skumanich-like magnetic braking law. The works of \cite{Sills2000} and \cite{Matt2015} find that $\Dot{J}\propto\Omega$ in the saturated regime.  \cite{2022MNRAS.517.4916E} \textcolor{black}{suggest that the multi-polar effect-dependent MB model of \cite{Garraffo2018} can also reproduce such scaling (however, see Fig.~\ref{fig:diffmod} and Section~\ref{subs:diffmodel}), and} that a similar scaling best reproduces observations of tidally locked close binaries. This is close to our relation for $p=4$ according to Eq.~(\ref{eq:jdotpropp4}). So we fix $p=4$ for the rest of the paper.

Figure~\ref{fig:p&fdz} also shows the behaviour of $f_\mathrm{DZ}$ with rotation and the evolution of the FCMD. The effect of varying $\Omega$ on the dead zone has been discussed in MS (see their section~2). Here we explore the behaviour of $f_\mathrm{DZ}$ in a star initially spinning up and contracting and later spinning down. As the star spins up, till $t\approx100\,\mathrm{Myr}$ (from the triangle to the circle in each subplot of Fig.~\ref{fig:p&fdz}), the dead zone increases ($f_\mathrm{DZ}$ decreases) because the combined effect of an increase in $\Omega$ and a decrease in $R_\ast$ increases the thermal and centrifugal pressure 
  as we move away from the surface. This balances magnetic pressure further out. From then onward the star begins to spin down but the behaviour of $f_\mathrm{DZ}$ does not stay the same throughout this time period. During the initial few $100\,\mathrm{Myr}$ $f_\mathrm{DZ}$ decreases as the spin decreases (from the circle to the star in each subplot of Fig.~\ref{fig:p&fdz}). As the star slows down, both the centrifugal pressure and magnetic pressure decrease close to the star, but the magnetic pressure falls much more steeply away from the star so the two balance out closer to the star and $f_\mathrm{DZ}$ increases (from the star to the plus in each subplot of Fig.~\ref{fig:p&fdz}). 

\subsection{Dependence on the initial spin and the disk-locking time-scale}
\label{subs:dependences}
\begin{figure}
\includegraphics[width=0.45\textwidth]{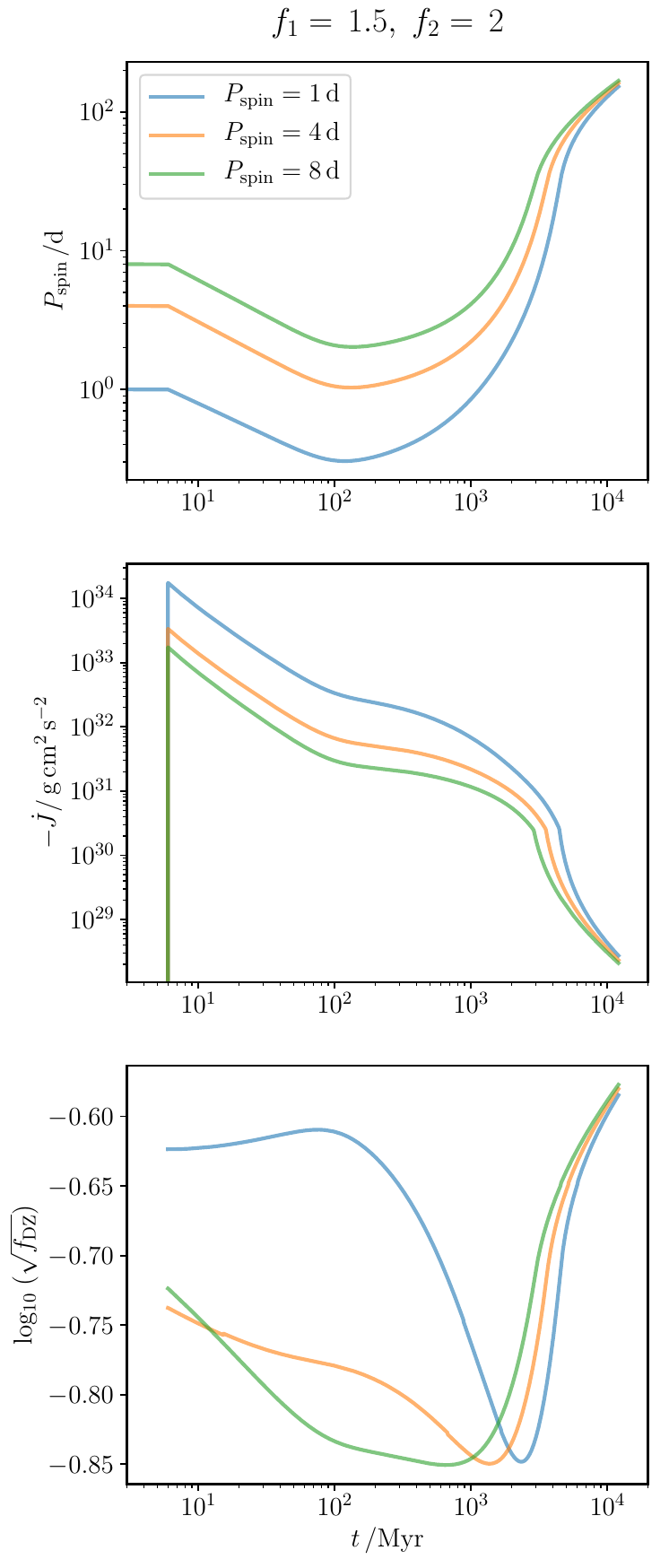}
\caption{The trajectories for the spin evolution of a FCMD with initial $P_\mathrm{spin}/\,\mathrm{d}\in \{1,\,4,\,8 \}$, $\tau_\mathrm{dl} = 6\,\mathrm{Myr}$, $p=4$ and the rest of the stellar parameters as for Fig.~\ref{fig:p&fdz}. The three subplots show the temporal evolution of $P_\mathrm{spin}$, the magnitude of the torque $\Dot{J}$ and $\sqrt{f_\mathrm{DZ}}$ (according to equation \ref{eq:jdotprop}).}
\label{fig:pini}
\end{figure}

In this section, we analyze the dependence of our model results on initial conditions $P_\mathrm{spin,0}$ and $\tau_\mathrm{dl}$. We first plot 
 {in Fig.~\ref{fig:pini}} 
the trajectories for a FCMD with initial $P_\mathrm{spin}/\,\mathrm{d}\in \{1,\,4,\,8 \}$, $\tau_\mathrm{dl} = 6\,\mathrm{Myr}$ and the rest of the stellar parameters the same as in Fig.~\ref{fig:p&fdz}. All systems spin up during their contraction phase and then spin down by MB. Systems that start at higher initial $P_\mathrm{spin}$ have larger $P_\mathrm{spin}$ at all times. The trajectories converge only when $t \gtrsim 5\,\mathrm{Gyr}$. It can be seen that the system with initial $P_\mathrm{spin} = 1\,\mathrm{d}$ experiences a greater spin-down torque throughout its evolution than those with larger initial periods. This is because a larger $\Omega$ leads to a stronger spin-down torque at all times (Eq.~(\ref{eq:jdot})). Interestingly, the system with an intermediate initial spin period has the smallest $f_\mathrm{DZ}$ for a few Myr. After this systems with higher $\Omega$ have larger $f_\mathrm{DZ}$ until a $t \approx 3$ Gyr. The trend then  {reverses} for the remainder of the evolution. This behaviour can be attributed to the rather complex interplay between the initial spin and the contraction over the first few Myr, followed by a non-trivial dependence of magnetic and centrifugal pressure on $\Omega$.

\begin{figure}
\includegraphics[width=0.45\textwidth]{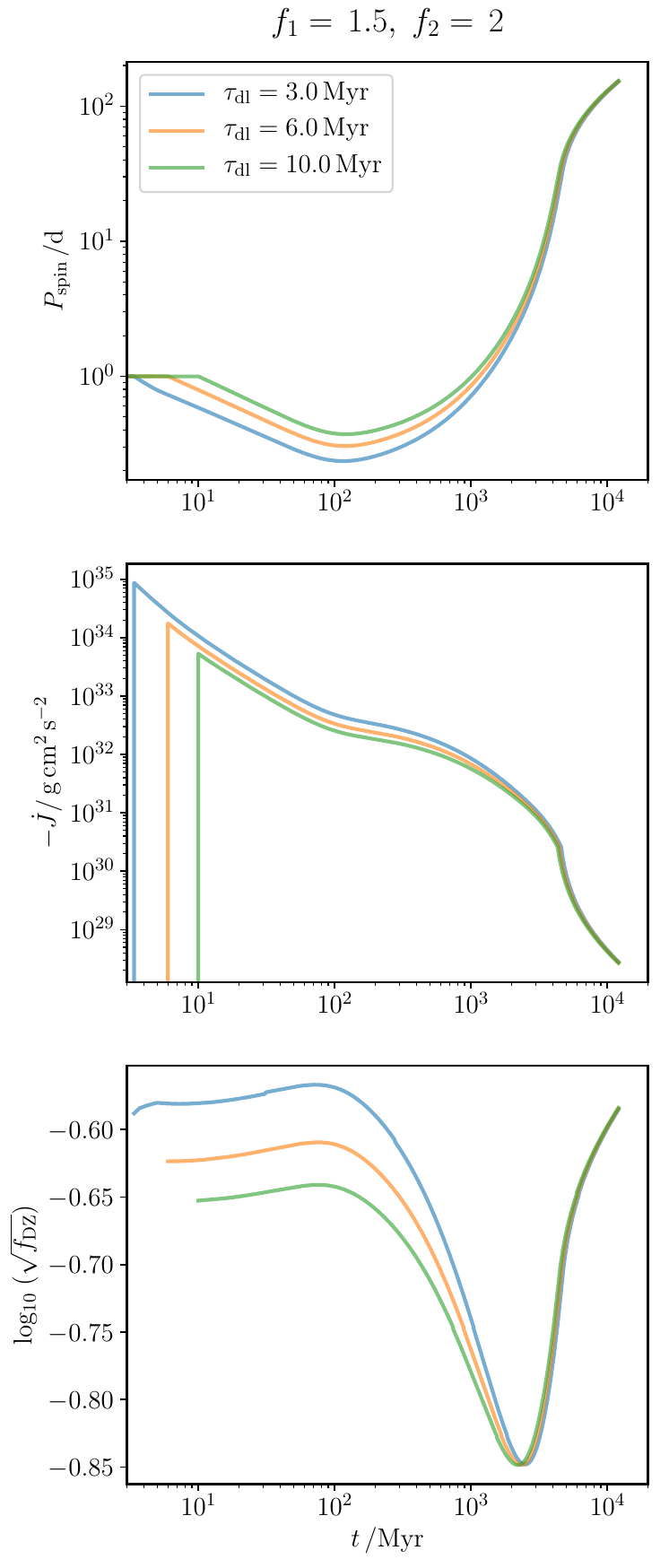}
\caption{The trajectories for the spin evolution of a FCMD with initial $P_\mathrm{spin} = 1\,\mathrm{d}$ and $\tau_\mathrm{dl}/\,\mathrm{Myr}\in \{3,\, 6,\, 10\}$, $p=4$ and the rest of the stellar parameters as for Fig.~\ref{fig:p&fdz}. The three subplots are as in Fig.~\ref{fig:pini}.}
\label{fig:tdl}
\end{figure}

We repeat this analysis, now fixing initial $P_\mathrm{spin} = 1\,\mathrm{d}$ and choosing $\tau_\mathrm{dl}/\,\mathrm{Myr}\in \{3,\, 6,\, 10\}$ for the same FCMD as in Fig.~\ref{fig:pini}. Our results are shown in Fig.~\ref{fig:tdl}.  The system whose disk disperses earlier evolves to smaller $P_\mathrm{spin}$ compared to the systems whose disks disperse later. This can be attributed to the fact that during the initial few Myr, the FCMD is a contracting PMS star. The star has a larger $R_\ast$ at earlier times and contracts to a unique $R_\ast$ by about 100 Myr. As a result, the PMS star that dispersed its disk earlier contracts  to a faster-spinning FCMD. In addition, this system experiences a stronger MB torque according to Eqs. (\ref{eq:ra}) and (\ref{eq:jdot}). Finally, the behaviour of $f_\mathrm{DZ}$ is qualitatively similar for each trajectory, with $f_\mathrm{DZ}$ relatively constant through the spin-up phase, followed by a steep decline and then increase at about 2 Gyr. This result shows that the spin evolution tracks depend on $\tau_\mathrm{dl}$ and trajectories only converge to the same $P_\mathrm{spin}$ after a few Gyr. Thus, we use $\tau_\mathrm{dl}$ as an initial condition as well as initial $P_\mathrm{spin}$. 
However, it is unlikely that these are independent of each other, as suggested by \cite{2018AJ....155..196R} who find that slow rotators possess long-lived disks and vice versa. In addition, different $\tau_\mathrm{dl}$ are expected to have varying relative significance, with observations showing a small fraction of stars still interacting with their disks after about 10 Myr \citep{2010A&A...510A..72F}. We also compare the MB torques of our FCMDs in their PMS phase to the results of \cite{Kounkel2023} and \cite{Gallet2019}. Fig.~\ref{fig:pini} shows that the $0.3\,M_\odot$ star with $P_\mathrm{spin} = 1\,\mathrm{d}$ and $\tau_\mathrm{dl} = 6\,\mathrm{Myr}$ experiences a torque of about $10^{34}\,$erg at the time of its disk dispersal, and Fig.~\ref{fig:pini} shows that the $0.3\,M_\odot$ star with $P_\mathrm{spin} = 1\,\mathrm{d}$ and $\tau_\mathrm{dl} = 3\,\mathrm{Myr}$ experiences a torque of about $10^{35}\,$erg at time of its disk dispersal. Our numbers are similar to that of \cite{Gallet2019}, and about an order of magnitude lower than that of \cite{Kounkel2023}. This may imply that additional sources of torque are at play during these early times which have not been considered in this work. A more rigorous study of MB at very early times in disk-locked PMS stars may be able to  {resolve} this problem \textcolor{black}{(e.g. \citealt{2019A&A...632A...6G,2020A&A...643A.129P,2021ApJ...906....4I,2021MNRAS.508.3710R})}. 

\subsection{The effect of dead zones}
\label{subs:deadzone}
\begin{figure*}
\centering
\includegraphics[width=\textwidth]{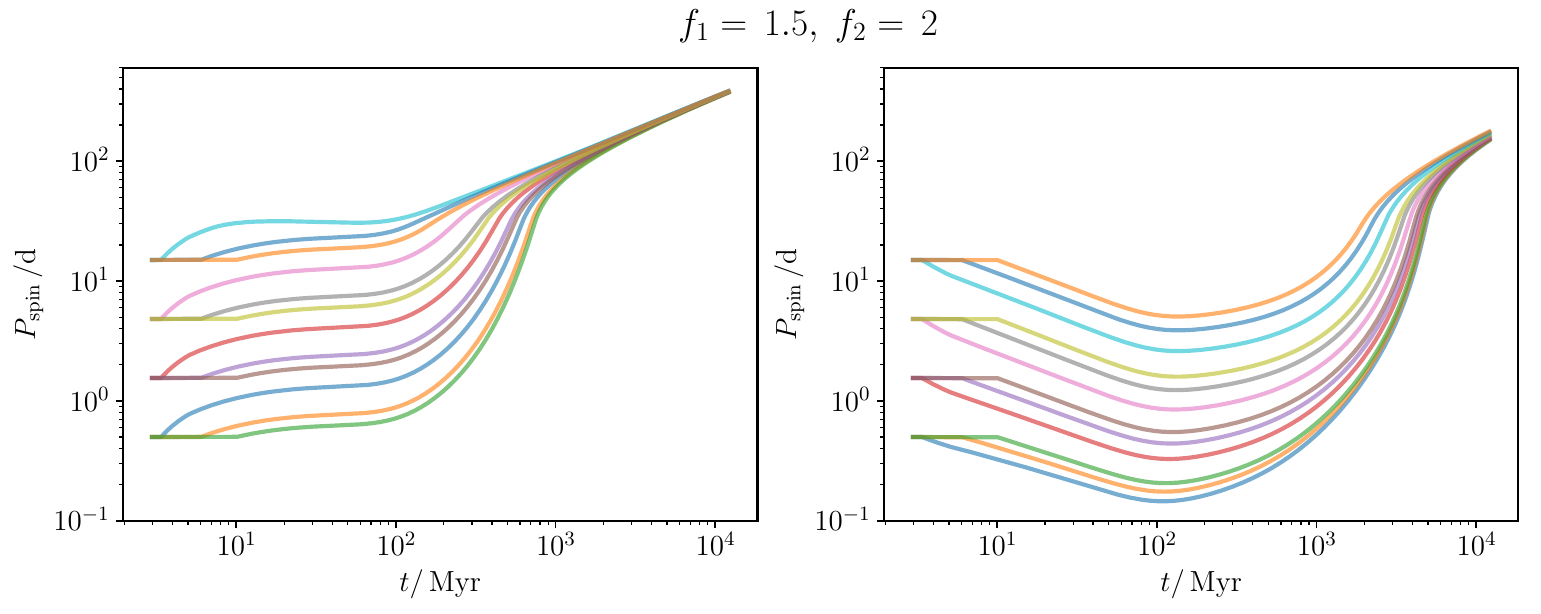}
\caption{Plots illustrating the dependence of the trajectories on the effect of the dead zone implementation. The trajectories show the spin evolution of a FCMD with the same stellar parameters as in Fig.~\ref{fig:p&fdz} with initial $P_\mathrm{spin}$ equally spaced between $[0.5,\,15]\,\mathrm{d}$ in log scale and $\tau_\mathrm{dl}/\,\mathrm{Myr} \in \{3,\,6,\,10\}$. The different colours denote different tuples $(P_\mathrm{spin},\,\tau_\mathrm{dl})$ with which a system begins its evolution. \textit{Left: }We implement no dead zone calculation such that $f_\mathrm{DZ}=1$.  \textit{Right: }We calculate $f_\mathrm{DZ}$ and implement it in the wind mass loss.}
\label{fig:bimod}
\end{figure*}

We illustrate the dependence of the trajectories on the effect of the dead zone implementation. \textcolor{black}{The modelling of dead zones has been done using simulations by works such as \cite{2015ApJ...798..116R,Garraffo2015}, where the latter showed that the winds are emitted, and as a consequence MB torque is exerted at mid-latitudes of the star. However, the exact latitude of the MB exertion is not shown to affect the overall rotational braking of the star. As a result, the effect of dead zones from simulation results has been incorporated in the MB torques of works such as \cite{Matt2015, Garraffo2018}. Our parameterized model also calculates an effective reduction in the overall wind mass loss rate of the star with $f_\mathrm{DZ}$. To show its effect on the spin-down, }
we evolve a FCMD with the same stellar parameters as in Fig.~\ref{fig:p&fdz} and making two sets of trajectories with initial $P_\mathrm{spin}$ equally spaced in log in the range $[0.5,\,15]\,\mathrm{d}$ and $\tau_\mathrm{dl}/\,\mathrm{Myr} \in \{3,\,6,\,10\}$. For one set of trajectories, we implement no dead zone calculation ($f_\mathrm{DZ}=1$). For the other set, we calculate $f_\mathrm{DZ}$ and implement it in the wind mass loss. The results of the evolution are shown in Fig.~\ref{fig:bimod}. 
For the case of $f_\mathrm{DZ}=1$, the MB torque during the initial phase of the evolution dominates over the spin-up owing to the contraction of the star, such that the stars never achieve a higher $\Omega$ than that with which they began their evolution. In other words, there is no spin-up phase for these systems. This effect is similar to what is seen in some of the trajectories of \citet[see their left subplot in Fig.~2]{Garraffo2018} \textcolor{black}{where they show for comparison, the evolution of systems with a Skumanich-type torque throughout, without accounting for the magnetic field complexity modulation.} Our trajectories for this case do show some spread in their spins until they converge to a single spin in about $700\,\mathrm{Myr}$. However, our trajectories for $f_\mathrm{DZ}=1$ are unable to explain the spread in spin periods of older OCs such as Praesepe \citep{2021ApJS..257...46G} or field stars \citep{Popinchalk2021}. On the other hand, the case where $f_\mathrm{DZ}$ is implemented shows clear epochs of spin up and spin down, with $P_\mathrm{spin}$ between about 0.1 d and 10 d at 100 Myr. The spread in spin periods persists till field ages (which we define to be times greater than about 1 Gyr) and all trajectories only converge by about 5 Gyr. In addition to it being a physical process that needs to be taken into account in model calculations, the inclusion of this phenomenon is able to well reproduce a spread in the spins of FCMDs till field ages. We also show, \textcolor{black}{in agreement with similar results to that of \cite{Amard2019,Gossage2021},} that the lifetime of this spread is mass-dependent, such that FCMDs with lower masses survive longer as rapid rotators when more massive FCMDs have already converged to become slow rotators (see Figs~\ref{fig:1.5&2}, \ref{fig:1.5&3}, \ref{fig:0.5&2}, \ref{fig:diffmod} and Section~\ref{subs:oc&field}).

\section{Results}
\label{sec:results}
In this section, we compare our results to observations with the goal of addressing the key observed features of FCMD spin properties. With a plethora of data available on  {FMCD in OCs and in the field,} we select a subset of them with good estimates of stellar mass, spin periods, and ages. We then attempt to estimate a few other observables the evolution of which can be tracked with our spin-down model.

\subsection{Comparison with spins of FMCDs in open clusters and 
in the field}
\label{subs:oc&field}
So far our model results depend on two free parameters. \textcolor{black}{The factor $f_1$ models the strength of the $\alpha$-effect, and consequently the strength of the poloidal field. The factor $f_2$ determines the stage where the convective viscosity is curtailed, and consequently where the system transitions from the saturated to the unsaturated regime.} Because these are very difficult to estimate, we illustrate the dependence of our model on $f_1$ and $f_2$. We assume for the remainder of this work that the same $f_1$ and $f_2$ govern the evolution of FCMDs of all masses. With this assumption, once we calibrate $f_1$ and $f_2$ to match with the observations of FCMDs of a given mass, our model is calibrated for FCMDs of all masses. For instance, owing to the dependence of $B_\mathrm{p}(R_\ast)\propto\gamma$ on stellar mass (Eq.~(\ref{eq:gamma})), the same $f_1$ yields a stronger magnetic field for heavier stars and vice versa (also see Section~\ref{subs:dynamo}).  We 
compute a set of trajectories with $M_\ast/\,M_\odot \in \{0.1,\,0.15,\,0.2,\,0.25,\,0.3,\,0.35\}$, initial $P_\mathrm{spin}$ equally spaced in log in the range $[0.2,\,35]\,\mathrm{d}$ and $\tau_\mathrm{dl}/\,\mathrm{Myr} \in\{3,\,10\}$. The minimum initial $P_\mathrm{spin}$ is motivated by the minimum initial $P_\mathrm{spin}$ in the trajectory of \citet[see their Fig.~2]{Garraffo2018} and \citet[see their Fig.~7]{Pass2022}.  The maximum initial $P_\mathrm{spin}$ is motivated by the maximum observed $P_\mathrm{spin}$ of FCMDs in the Orion Nebula Cluster (ONC, \citealt{1997AJ....113.1733H}), where the youngest stellar population may be approximately 1 Myr old \citep{2019A&A...627A..57J}. 

\begin{figure*}
\centering
\includegraphics[width=0.95\textwidth]{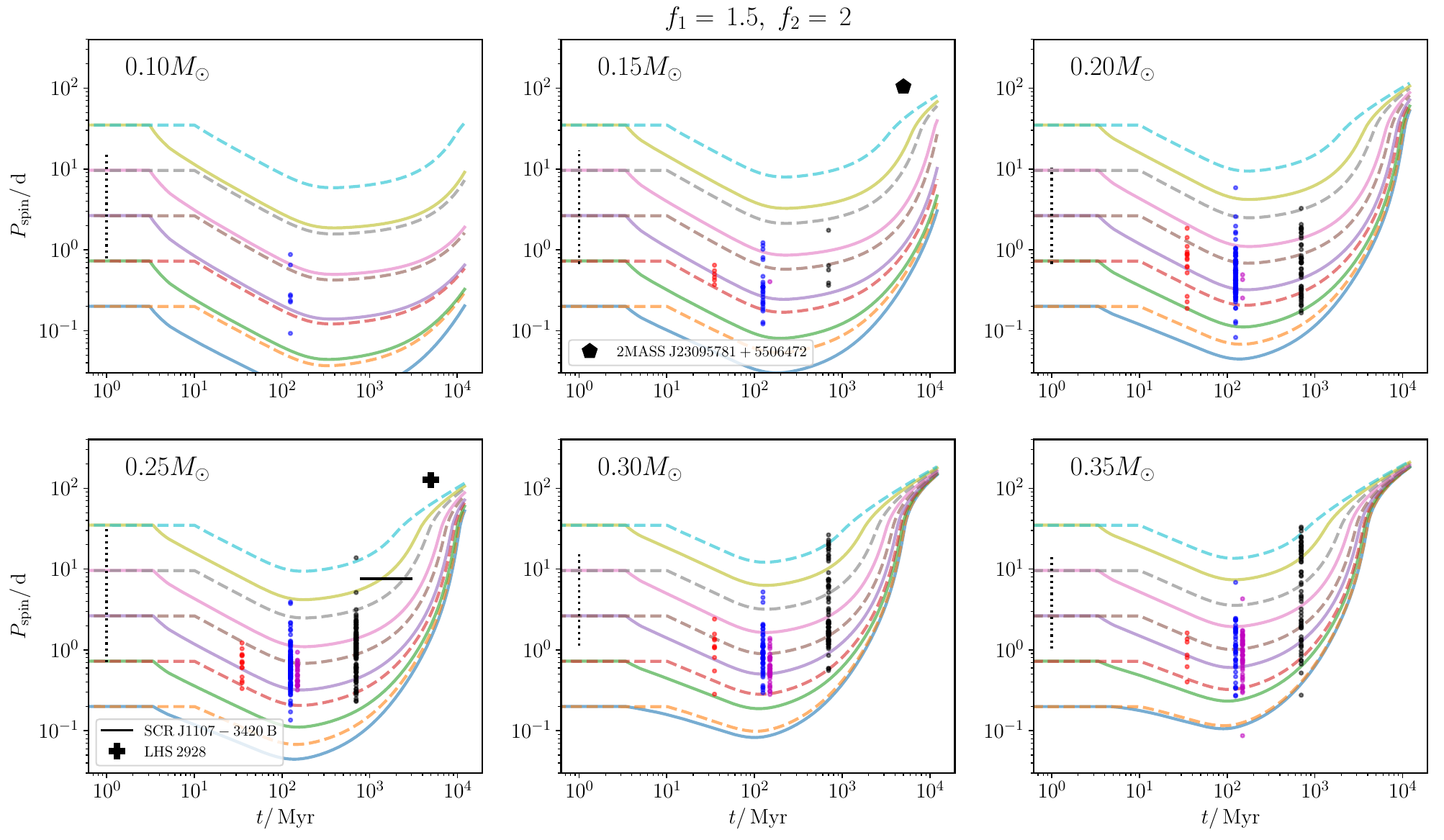}
\caption{ The spin evolution 
along the trajectories  {of stars} 
with $M_\ast/\,M_\odot \in \{0.1,\,0.15,\,0.2,\,0.25,\,0.3,\,0.35\}$, initial $P_\mathrm{spin}$ equally spaced in log in the range $[0.2,\,35]\,\mathrm{d}$, and $\tau_\mathrm{dl}/\,\mathrm{Myr} \in\{3,\,10\}$ (solid and dotted lines respectively) for $f_1=1.5$ and $f_2=2$. The observations of OCs   {stars} from \protect\cite{2021ApJS..257...46G} are plotted as a vertical array of dots, such that the dots in red are systems from NGC2547 (about 35 Myr old), dots in blue are Pleiades (about 125 Myr old), dots in magenta are NGC2516 (about 150 Myr old), and dots in black are Praesepe (about 700 Myr old). The dotted line at 1 Myr spans the maximum and minimum spins observed in each mass bin in the ONC by \protect\cite{1997AJ....113.1733H}. $\mathrm{2MASS\,J23095781\,+\,5506472}$ and $\mathrm{LHS \,2928}$ with their minimum inferred ages, $\mathrm{SCR\, J1107\, -\, 3420\,B}$ with its age range and spin periods \citep{Pass2022} are also plotted. }
\label{fig:1.5&2}
\end{figure*}

\begin{figure*}
\centering
\includegraphics[width=0.95\textwidth]{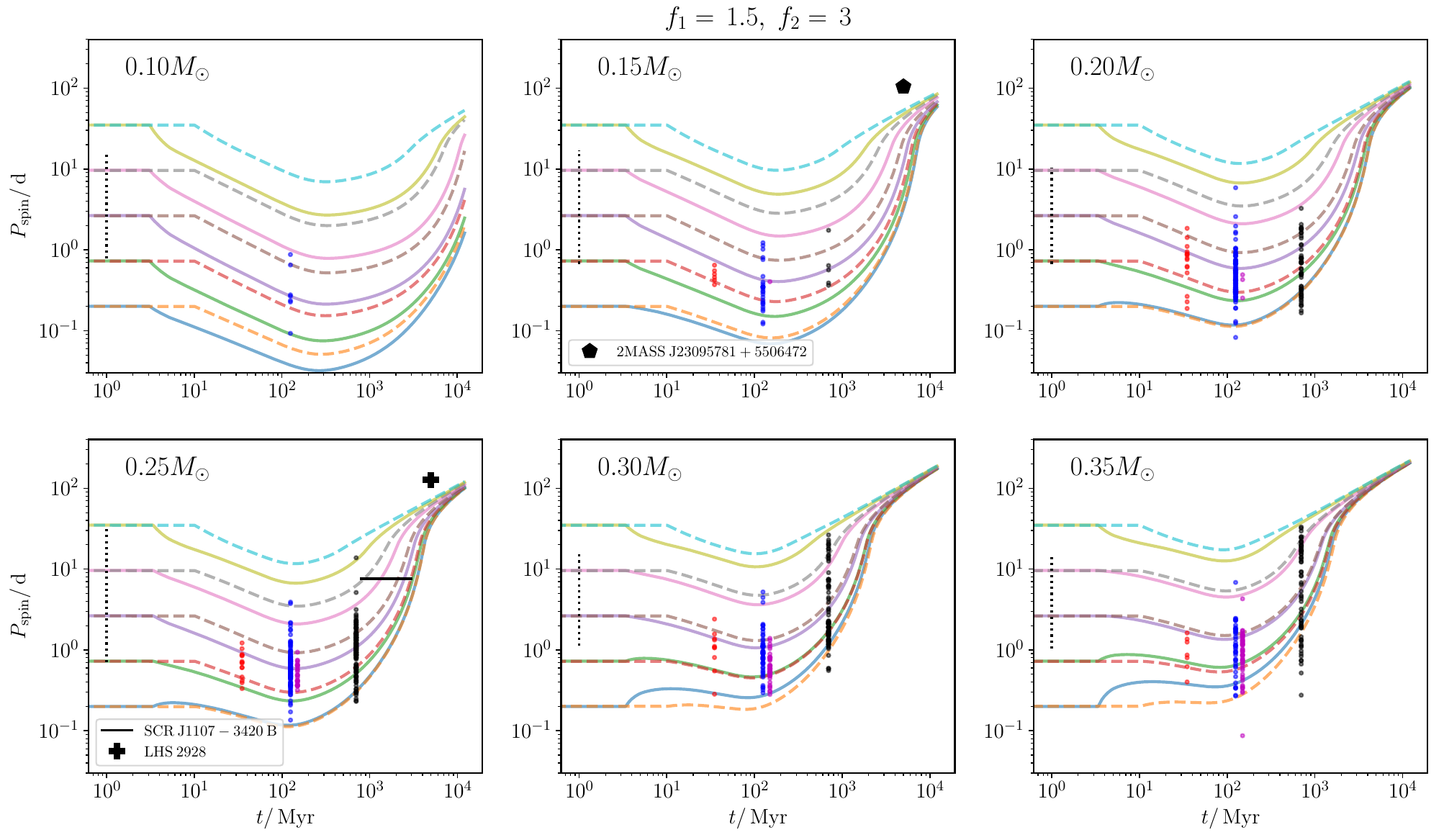}
\caption{The same plot as Fig.~\ref{fig:1.5&2} but for $f_1=1.5$ and $f_2=3$.}
\label{fig:1.5&3}
\end{figure*}

\begin{figure*}
\centering
\includegraphics[width=0.95\textwidth]{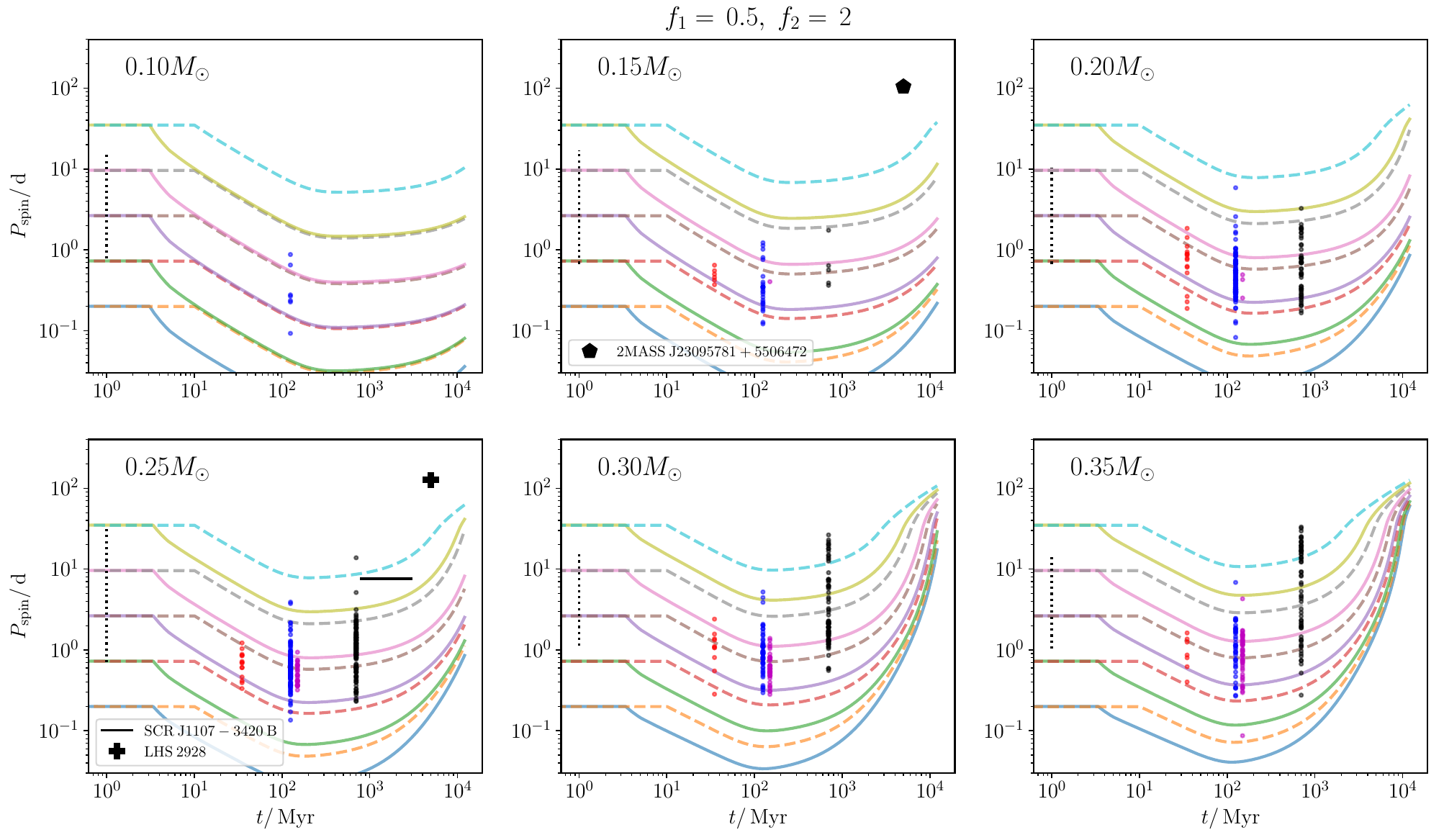}
\caption{The same plot as Fig.~\ref{fig:1.5&2} but for $f_1=0.5$ and $f_2=2$.}
\label{fig:0.5&2}
\end{figure*}

The two $\tau_\mathrm{dl}$ are chosen as the upper and lower bounds to the time-scales over which contracting PMS stars end their accretion phase \citep{2010A&A...510A..72F}. From equations (\ref{eq:mlconv}) to (\ref{eq:jdot}) we can write
\begin{center}
\begin{equation}
\label{eq:jdotpropf}
\Dot{J} \propto
    \begin{cases}
        {f_2}^{\frac{p}{2}}f_1, & \text{when   }\; 2\pi f_2/\tau_\mathrm{c}\Omega \; \leq \; 1,\\
        \\
        {f_1}, & \text{otherwise.}
    \end{cases}
\end{equation}
\end{center}
From Eq.~(\ref{eq:jdotpropf}) we  can see that both $f_1$ and $f_2$ influence the strength of the MB torque, while $f_2$ also governs when the system transitions from the saturated to the unsaturated regime. Our results for three sets of choices for $f_1$ and $f_2$ are shown in Figs~\ref{fig:1.5&2}, \ref{fig:1.5&3} and \ref{fig:0.5&2}. 
Across the figures, we see that larger $f_1$ and $f_2$ lead to stronger MB torques and, consequently, similar tracks have larger periods at all times. The figures also illustrate the mass dependence of the spread and the eventual convergence of the spins of the systems, where FCMDs with larger masses converge earlier and spin down more than lower-mass FCMDs. We also plot stellar spins from OCs observed by \citet[hereinafter G21]{2021ApJS..257...46G}. In these clusters, potential contamination by field stars has been removed. We consider all FCMDs with masses in the range $[0.075,\,0.375)\,M_\odot$ in each OC dataset of G21. We then plot the spin period of each FCMD (table~F1 of G21) against their inferred age (table~2 of G21) and assign it to one of the six subplots $M_\mathrm{FCMD}/\,M_\odot \in \{0.1,\,0.15,\,0.2,\,0.25,\,0.3,\,0.35\}$ if the mass of the FCMD lies between $[M_\mathrm{FCMD}-0.025,\,M_\mathrm{FCMD}+0.025)\,M_\odot$. The observations are plotted as a vertical array of dots, such that the dots in red are systems from NGC2547 (about 35 Myr old), blue are Pleiades (about 125 Myr old), magenta are NGC2516 (about 150 Myr old), and black are Praesepe (about 700 Myr old). Table~2 of G21 also shows the deviation of the metallicity of these OCs from solar. 
We find that varying the metallicity of our 
FCMDs does not affect the spin-down history of our trajectories (Fig.~\ref{fig:Z}) and so we only work with solar metallicity FCMDs. We also illustrate the likely initial spins that  {stars in} an OC may have evolved with, by plotting the maximum and minimum spin period observed in each mass bin of the ONC from the data of \cite{1997AJ....113.1733H}. This is shown as a thin dotted vertical line at 1 Myr.
We see that the most favourable match of OC data to our trajectories is in Fig.~\ref{fig:1.5&2} where, for each mass bin, if we begin with a similar range of spins as in the ONC, our model can reproduce the entire observed spread of spins in  NGC2547, NGC2516 and the Pleiades. An initial range of spins from the ONC can explain the entire observed range of spins in Praesepe except for the $0.3M_\odot$ and $0.35M_\odot$ subplots. We propose that in order to explain with our model a population of stars in Praesepe with $P_\mathrm{spin}\gtrsim20\,\mathrm{d}$ and masses $M$, $0.30 \leq M/\,M_\odot \leq 0.35$, the initial spin distribution of Praesepe ought to have had a population of slow rotators of such masses and an initial $P_\mathrm{spin}\approx30\,\mathrm{d}$. We note that \cite{2019A&A...627A..57J} suggest that ONC harbours stellar populations with different ages, with the youngest about 1.4 Myr old and the oldest about 4.5 Myr old. So in its oldest population, a number of stars may have already dispersed from their circumstellar disks and begun their spin evolution. It would be intriguing to find whether under the same assumptions on the $\tau_{\rm dl}$ 
it will be possible to reproduce spin distributions in particular populations of ONC. 

We also plot FCMDs with ages greater than about 1 Gyr (which we define as field stars), but owing to poor estimates of their inferred ages we only plot a few of them from the data of \citet{Pass2022}.
We choose field stars with masses and spin periods from their Table~3 and inferred ages from their Sec.~3. We take the minimum age of 2MASS J23095781+5506472 to be 4.7 Gyr, the minimum age of LHS 2928 to be 5.9 Gyr and an age range of SCR J1107-3420 B between 0.8 to 3 Gyr. We find that our stars are modestly rapidly rotating to match well with 2MASS J23095781+5506472 and LHS 2928 in Fig.~\ref{fig:1.5&2}, with a good match with SCR J1107-3420 B. The match is better for field stars in Fig.~\ref{fig:1.5&3}. However, we note that the inferred ages we plot here are minima (\citealt{Pass2022}) and a better match with our trajectories can be obtained in Fig.~\ref{fig:1.5&2} if the true ages of 2MASS J23095781+5506472 and LHS 2928 are a few Gyr more than their inferred minima.

\subsection{Comparison with other magnetic braking models}
\label{subs:diffmodel}
\begin{figure*}
\centering
\includegraphics[width=0.95\textwidth]{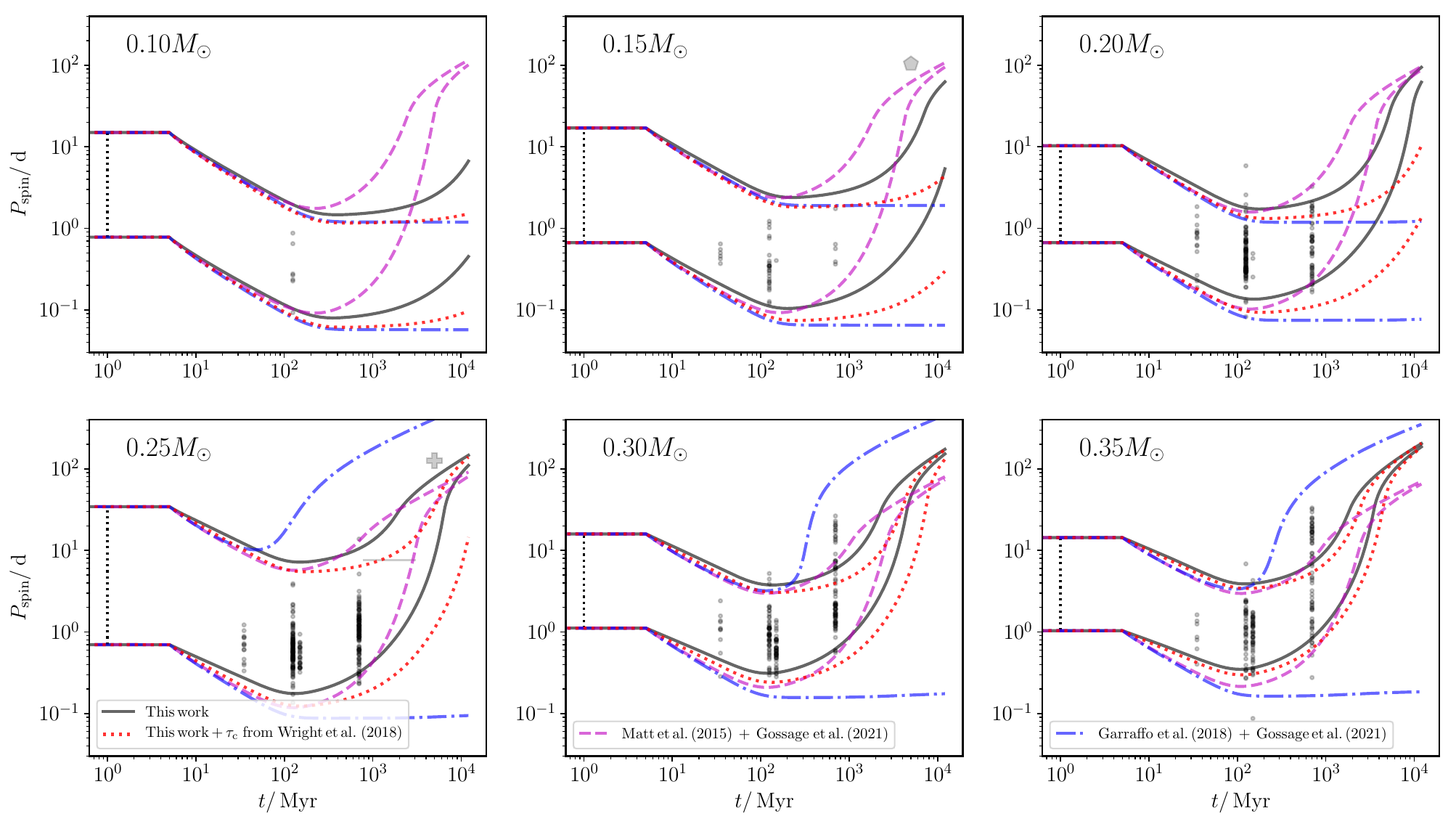}
\caption{ The spin evolution 
along the trajectories  of stars
with $M_\ast/\,M_\odot \in \{0.1,\,0.15,\,0.2,\,0.25,\,0.3,\,0.35\}$ with initial spins as the maximum and minimum spins observed in each mass bin in the ONC (\protect\citealt{1997AJ....113.1733H}, shown as a vertical dotted line at 1 Myr), and $\tau_\mathrm{dl}= 5\,\mathrm{Myr}$ for our MB model with $f_1=1.5$ and $f_2=2$ (solid black lines), for the MB model by \protect\cite{Matt2015} calibrated by \protect\cite{Gossage2021} (dashed magenta lines), for the MB model by \protect\cite{Garraffo2018} calibrated by \protect\cite{Gossage2021} (dash-dotted blue lines) and our model with $f_1=1.5$ and $f_2=2$ but with $\tau_\mathrm{c}$ from \protect\cite{2018MNRAS.479.2351W} (dotted red lines). For the models from other works $\tau_\mathrm{c}$ is taken from \protect\cite{2018MNRAS.479.2351W}. The observations of OCs shown as grey dots are the same as in Fig.~\ref{fig:1.5&2}. }
\label{fig:diffmod}
\end{figure*}

\begin{figure}
\includegraphics[width=0.45\textwidth]{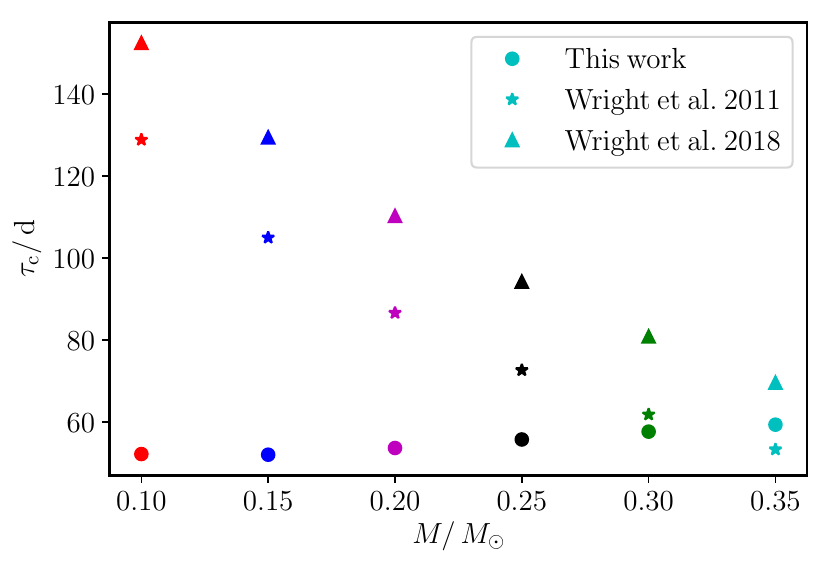}
\caption{The comparison of the estimates of $\tau_\mathrm{c}$ from this work with the estimates from  \citet{2011ApJ...743...48W} and \citet{2018MNRAS.479.2351W}. Owing to the time-dependence of our $\tau_\mathrm{c}$ estimates, we plot $\tau_\mathrm{c}$ from our work for FCMDs beyond their contraction phase when our $\tau_\mathrm{c}$ attains an approximately constant value. }
\label{fig:tauc}
\end{figure}

\textcolor{black}{We now compare the spin-down results of our model to that of other MB models. First, we use the model by \cite{Matt2015} in which the stellar torque is a function of the Rossby number $R_\mathrm{o}= 1/\,(\Omega\,\tau_\mathrm{c})$. Their model relates uncertain stellar parameters such as magnetic fields and winds to a simple physically motivated relation (see their equations 4 and 5). Upon scaling the strength of the torque empirically, they obtain a bifurcated MB torque wherein $\Dot{J}\propto\Omega$ in the saturated regime and $\Dot{J}\propto\Omega^3$ in the unsaturated regime (see their equations 6-8). The transition from the saturated to the unsaturated regime is when $R_\mathrm{o}$ becomes greater than a critical value (see their equation 2). Although derived by different means, their MB torque looks functionally similar to our equation (\ref{eq:jdotpropp4}). This MB model has been calibrated in recent works with different estimates of $\tau_\mathrm{c}$ (see for instance \citealt{Amard2019}, \citealt{Gossage2021}). We use the calibration by \citet[their table~1 and equations 2-4]{Gossage2021} with $\tau_\mathrm{c}$ from \cite{2018MNRAS.479.2351W}. Secondly, we use the MB model from \cite{Garraffo2018}. They derive a magnetic-morphology-driven spin-down torque based on simulation results by \cite{Garraffo2015,Garraffo2016}. They construct a MB model wherein the spin-down torque is curtailed if higher-order multipolar effects are at play, by adding a complexity modulation factor to an otherwise Skumanich-like MB torque. We use its recent calibration by \citet[their table~2 and equations 6-10]{Gossage2021} with $\tau_\mathrm{c}$ from \cite{2018MNRAS.479.2351W}. Finally, we analyse the effect of varying $\tau_\mathrm{c}$ in our model by using the estimates of \citet[see their equation 6]{2018MNRAS.479.2351W} instead of deriving $\tau_\mathrm{c}$ using the mixing-length theory (MLT). The results are shown in Fig.~\ref{fig:diffmod}. Till about 1 Gyr, the parameter space covered by our model and the modified model of \cite{Matt2015} is similar. However, beyond a Gyr the modified model of \cite{Matt2015} leads to even the least massive FCMDs to transition from the saturated to the unsaturated regime within the Galactic age. This is because the calibration of \cite{Gossage2021} leads to a critical $R_\mathrm{o}\approx 0.02$ which is smaller than that in our model which is about 0.07, so the system reaches its critical $R_\mathrm{o}$ earlier. However, once the system becomes unsaturated, our model leads to a stronger torque and leads to more massive FCMDs to have $P_\mathrm{spin}\gtrsim100\,$d by 10 Gyr. Comparing with the field stars in our plots, we see that the model of \cite{Matt2015} agrees better with slow-spinning FCMDs less massive than about $0.2M_\odot$ while our model agrees better with slow-spinning massive FCMDs. However, our model is better at reproducing the spread in spin periods of late M-dwarfs in fig.~12 of \cite{Popinchalk2021}. More robust data on low-mass FCMDs with field ages will help us better constrain the behaviour of these models at field ages. The model by \cite{Garraffo2018} leads to weak MB for FCMDs less massive than about $0.2M_\odot$ such that the systems maintain an almost constant $P_\mathrm{spin}$ after the PMS contraction phase. For heavier FCMDs, the trajecetory with a smaller initial spin period experiences a weak MB and the systems behave similar to low-mass FCMDs. However, the system with a larger initial spin period experiences a strong MB after the PMS contraction and quickly spins down by about 100 Myr. This behaviour has also been seen in the trajectories of FCMDs plotted by \cite{Pass2022} with the \cite{Garraffo2018} model (see their fig.~7), and that of solar-like stars plotted by \citet[see their fig.~6]{Gossage2021}. Overall, the MB model by \cite{Garraffo2018} does not yield a good match with observations of low-mass FCMDs and yields two qualitatively distinct trajecetories for heavier FCMDs. The possibility that multi-polar effects do not affect MB efficiency has been explored by \cite{2019ApJ...886..120S}, but it is possible that these effects influence the evolution of systems at later stages of evolution \citep{2019ApJ...886..120S,vanSaders2019}. However, in the context of FCMDs which remain at main-sequence till the Galactic age, we attest that a MB torque dependent on multi-polar effects seems to be in tension with observations as well as other MB model trajectories.
We see that changing $\tau_\mathrm{c}$ in our model leads to significant changes in the spin evolution of low-mass FCMDs with the changes becoming negligible for more massive FCMDs. This is evident in Fig.~\ref{fig:tauc} where we compare our $\tau_\mathrm{c}$ estimates to that of \cite{2018MNRAS.479.2351W} and an earlier estimate by \cite{2011ApJ...743...48W}. Since our $\tau_\mathrm{c}$ is a function of stellar parameters, this imparts an additional time-dependence to it during the pre-main-sequence contraction phase, which is absent in the empirical fits. We see that our estimates predict lower values of $\tau_\mathrm{c}$ than the empirical fits. Our $\tau_\mathrm{c}$ is in better agreement with empirical fits for more massive FCMDs, thereby explaining the difference in the spin evolution for low-mass FCMDs and negligible difference for heavier FCMDs.}

\subsection{Comparison with other observables}
\label{subs:obs}

\begin{figure*}
\centering
\includegraphics[width=0.85\textwidth]{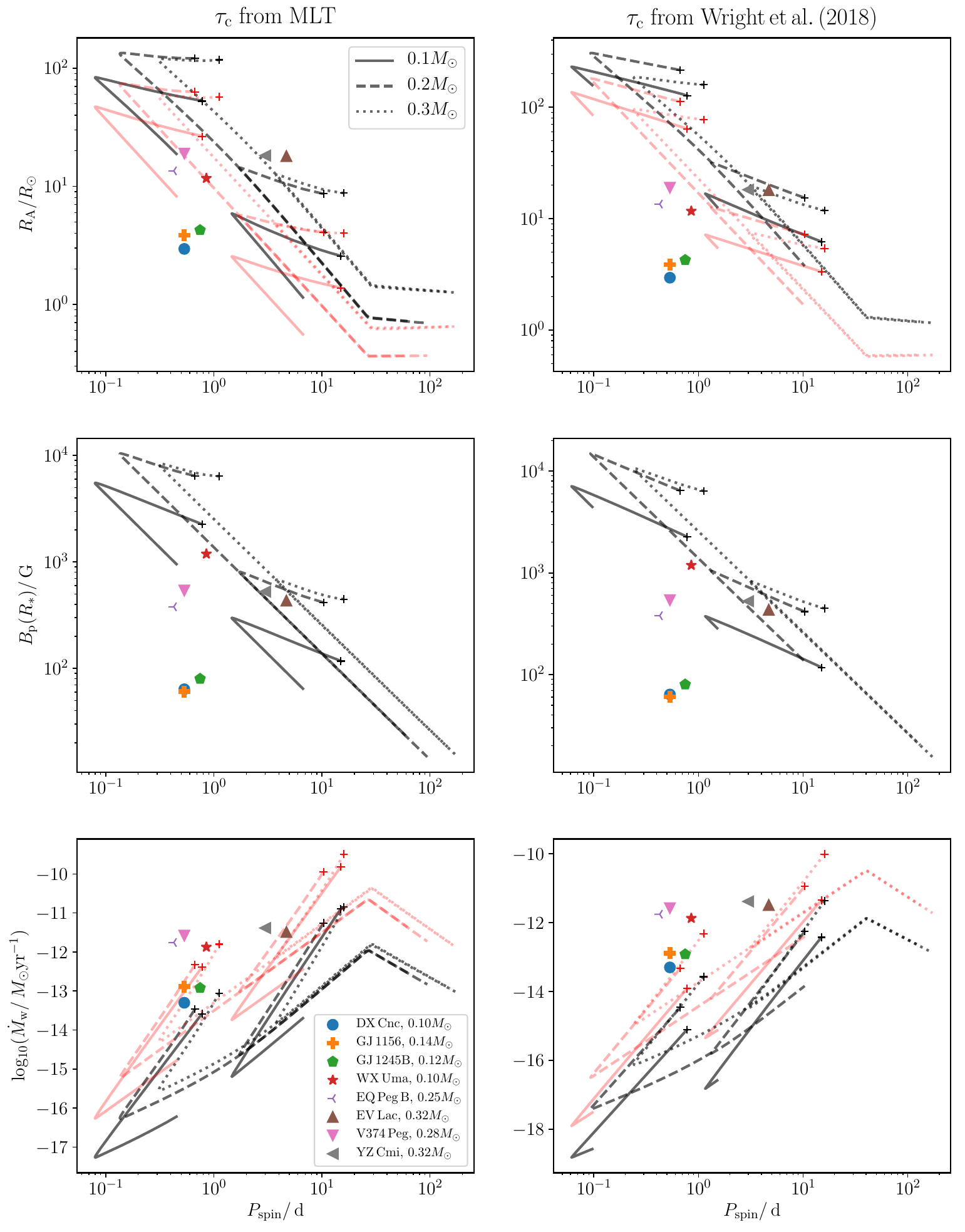}
\caption{The evolution of Alfvén radius $R_\mathrm{A}$, surface poloidal field $B_\mathrm{p}(R_\ast)$, and wind mass loss $\Dot{M}_\mathrm{w}$ as a function of $P_\mathrm{spin}$ for a set of trajectories for FCMDs with $M_\ast/\,M_\odot \in \{0.1,\,\,0.2,\,0.3\}$, initial spins as the maximum and minimum spins observed in each mass bin in the ONC by \protect\cite{1997AJ....113.1733H} and $\tau_\mathrm{dl}= 5\,\mathrm{Myr}$ for two sets of trajectories, one with $\tau_\mathrm{c}$ from the MLT (left column) and the other with $\tau_\mathrm{c}$ estimated by \protect\cite{2018MNRAS.479.2351W} (right column). The different line styles indicate FCMDs of different masses. Trajectories in red are the same trajecories in black, but $\Dot{M}_\mathrm{w}$ is scaled by $f_\mathrm{DZ}^{-1}$ and $R_\mathrm{A}$ by $f_\mathrm{DZ}^{1/4}$. The starting points of our trajectories are shown as pluses. Observations of FCMD parameters are those of \protect\cite{2018MNRAS.475L..25V}. }
\label{fig:obs}
\end{figure*}

\textcolor{black}{An advantage of our} spin-down model \textcolor{black}{is that it relates the properties of the stellar dynamo to physically motivated estimations of the magnetic field strength and stellar wind, thus} allowing us to track the evolution of observationally inferred parameters such as wind mass loss, surface magnetic fields and Alfvén radii of FCMDs using Eqs. (\ref{eq:mlconv}), (\ref{eq:bpconv}) and (\ref{eq:ra}). We compare these parameters with observations of \citet[Table~1]{2018MNRAS.475L..25V} and \cite{See2016} in Fig.~\ref{fig:obs}. We plot a set of trajectories of FCMDs with $M_\ast/\,M_\odot \in \{0.1,\,\,0.2,\,0.3\}$, initial spins as the maximum and minimum spins observed in each mass bin in the ONC by \cite{1997AJ....113.1733H} and $\tau_\mathrm{dl}= 5\,\mathrm{Myr}$. \textcolor{black}{Similar to Section~\ref{subs:diffmodel}, we plot two sets of such trajectories, one with $\tau_\mathrm{c}$ from MLT and the other with $\tau_\mathrm{c}$ estimated by \cite{2018MNRAS.479.2351W}. These trajectories are shown in black.} Upon plotting our trajectories as a function of $P_\mathrm{spin}$ we see that, after the initial spin-up owing to contraction, $B_\mathrm{p}(R_\ast)$ and $R_\mathrm{A}$ follow a simple power-law relation to $P_\mathrm{spin}$, as expected from Eqs. (\ref{eq:bpconv}) and (\ref{eq:ra}), until the FCMD transitions to the unsaturated regime causing a change in $\Dot{M}_\mathrm{w}$ and consequently $R_\mathrm{A}$. There is a deviation from a power-law behaviour of $\Dot{M}_\mathrm{w}$ with $P_\mathrm{spin}$ owing to the difference in the behaviour of $f_\mathrm{DZ}$ in different regimes of the spin-down process (the bottom two subplots in Fig.~\ref{fig:p&fdz}). The effect of $f_\mathrm{DZ}$ is suppressed in the evolution of the Alfvén radius because $R_\mathrm{A}\propto f_\mathrm{DZ}^{-1/4}$. \textcolor{black}{Changing the estimation of $\tau_\mathrm{c}$ from the MLT to that of \cite{2018MNRAS.479.2351W} leads to about an order of magnitude lower wind mass loss, and consequently an increase in the Alfvén radius for the $0.1M_\odot$ FCMD. This difference is negligible for the heavier FCMDs. This is because our $\tau_\mathrm{c}$ estimates agree better for heavier FCMDs (Figs~\ref{fig:diffmod} and \ref{fig:tauc}).}
We find that the Alfvén radii and poloidal fields agree with the observed estimates within a factor of 10 at worst (DX Cnc, GJ1156, and GJ 1245B). The discrepancy in $B_\mathrm{p}(R_\ast)$ can either be the result of assumptions by \cite{See2016}, such as a constant source radius which ought to be dependent on stellar parameters (their Sec.~2), or can be the result of a more fundamental process at play in FCMDs, such as strong-field and weak-field dynamos proposed by \cite{2011MNRAS.418L.133M}. Our wind mass loss estimates are lower by about one to two orders of magnitude than the observed estimates for FCMDs. These rates are difficult to measure accurately, with discrepancies even between empirical-model results and MHD simulation results for solar-like stars (see fig.~5 of \citealt{See2016} \textcolor{black}{where 3D MHD simulation results fail to estimate observationally inferred results at lower Rossby numbers).} If dead zones are at play in such stars, it will lead to a non-spherically symmetric wind mass-loss profile, wherein the poles will have stronger winds while the equators will have virtually no winds. \textcolor{black}{Although they scale their wind density according to the stellar activity, }the fact that \cite{See2016} have inferred the wind mass loss values assuming spherical symmetry may bring about an additional possibility that they have overestimated these values. \textcolor{black}{To test this we plot, for the same set of trajectories, new trajectories wherein we show the extent to which the observables can be overestimated, by scaling our modelled $\Dot{M}_\mathrm{w}$ by $f_\mathrm{DZ}^{-1}$ and $R_\mathrm{A}$ by $f_\mathrm{DZ}^{1/4}$ (equations~\ref{eq:mlconv} and \ref{eq:ra}). These trajectories are shown in red. We see that these trajectories agree better with the observations in general, illustrating that there may indeed be some overestimation in inferring these observations owing to the assumption of spherical symmetry by \cite{See2016}.}
Despite these differences between our approach and that of \cite{See2016}, it is encouraging to see that the results \textcolor{black}{from our parametrized model} are  {at least} close to the inferred observational parameters of \cite{2018MNRAS.475L..25V}.

\section{Discussion}
\label{sec:discussion}
In this section, we discuss the implications of the results of this paper and the work that we plan to undertake in the near future.
\subsection{Population synthesis}
\label{subs:popsynth}
With our spin-down model, we have been able to address the spread in the observed spins of FCMDs in OCs as well as field stars. 
In our analysis we use the {term } ``spread'', rather than ``bimodality'' {for } the spins of OCs, reported by \cite{2003ApJ...586..464B} who find the existence of fast- and slow-rotators (labeled C and I sequences), but relatively fewer intermediate-rotators. Whether or not our model can explain this inferred bimodality can be answered with population synthesis calculations, similar to the work of \cite{Brown2014} who used a set of initial spin distribution and disk-locking time. These are likely correlated (see Sec.~\ref{subs:dependences}). \textcolor{black}{Since our spin-down torque can be simply evaluated using stellar luminosity, stellar radius and
stellar mass, doing population synthesis studies with this model is relatively easy.} We shall undertake such a population synthesis study in future work. Our motivation now was to construct a simple working model of the spin-down torque and compare it with observations and other models. However, we note that this bimodality is not apparent for FCMDs in the recent observational dataset of G21 (see their Figs~10 and 12 and also our Fig.~\ref{fig:1.5&2}), where it appears that there are relatively abundant intermediate-rotators in some OCs.

\subsection{Extension to solar-type stars}
\label{subs:solarlike}
With promising results from this work, we shall in the future extend our spin-down model to explain solar-like stars, $0.35M_\odot \lesssim M_\ast\lesssim1.4M_\odot$, with convective envelopes and radiative cores  with a double-dynamo-like MB model as discussed by \cite{2022MNRAS.513.4169S} for CVs. The core being radiative is bound to induce additional complexities as well as extra stellar parameters to model the spin evolution for such stars. This complexity, to some extent, exists in our current FCMD models as well. Our models show that FCMDs with $0.30M_\odot \lesssim M_\ast\lesssim0.35M_\odot$ briefly develop a radiative core for a few Myr before it disappears. The mass of this core and  duration of its existence scales as the mass of the FCMD. We have neglected this effect because we observed that our evolutionary tracks only showed minor deviations from the assumption of the FCMDs being fully convective throughout their lifetime. However, this radiative core will persist for more massive M dwarfs, the effect of which needs to be properly accounted for in the future. The results of recent works such as \cite{Btrisey2023} can provide a better understanding of the transport of angular momentum in radiative regions, which can help explain the results of \cite{jao2023mind} who find that more massive FCMDs lose angular momentum faster than both less massive FCMDs as well as partly convective M dwarfs. Moreover, the core may have a radial dependence on $\Omega$, as seen in the solar interior models of \citet[their Fig.~2]{Eggenberger2019}. The complications can possibly be relieved with more robust data on MB in solar-like stars \citep{2023ApJ...948L...6M, getman2023magnetic}, as well as data {on the} Sun.

\subsection{Implications for star-planet interaction}
\label{subs:life}
While M dwarfs have an extremely long-lasting stable main-sequence, the vicinity of any exoplanets orbiting (say, the habitable region) around  active M dwarfs is affected by the extreme stellar environment owing to their close proximity to the host star. The atmospheres of these exoplanets can be highly irradiated, or in some cases completely stripped by the intense magnetized stellar winds. {For instance} \cite{Garraffo2017} and \cite{Cohen2018} have studied this effect on the TRAPPIST-1 system \citep{Gillon2017}. With our model, we obtain time-dependent analytical estimates of wind velocities, magnetic fields, and Alfvén radii in the vicinity of FCMDs which can be extended to obtain a first-order estimate of these parameters in the vicinity of the orbiting exoplanets. This can then be used to quantify the effect of the host star on the planetary atmosphere (similar to section~2 of \citealt{Cohen2018}, who used the method described by \citealt{2008JGRA..113.5214K}).

\section{Conclusions}
\label{sec:conclusions}
We have constructed a magnetic braking model with the goal of addressing the salient features of the spin-down of fully convective M dwarf (FCMD) stars. This angular momentum-loss model was first introduced for contracting pre-main-sequence stars by \cite{1992MNRAS.256..269T} and also used to explain the orbital period characteristics of cataclysmic variables \citep{Zangrilli1997} with fully convective donors. In this model magnetic braking happens because of an $\alpha-\Omega$ dynamo operating in the FCMD.  {In this work} we have improved upon our model using observations of the scaling of the magnetic braking torque with spin (\citealt{2011ApJ...743...48W} and subsequent works), as well as the process of entrapment of gas in dead zones, owing to an interplay of magnetic, thermal and centrifugal pressures (\citealt{1987MNRAS.226...57M} and Section~\ref{subs:deadzone}). The construction of our model depends only on the time evolution of stellar luminosity, stellar radius and stellar mass, \textcolor{black}{and is dependent on two free parameters; $f_1$ which models the strength of the poloidal field and $f_2$ which denotes the spin-down stage where the system transitions from the saturated to the unsaturated regime. These free parameters can be easily updated in light of more robust data in the future.} 

We find that \textcolor{black}{in accordance with previous works,} the spin evolution of our FCMDs depends not only on the initial spin of the star but also on the lifetime of any circumstellar disk around the star (Section~\ref{subs:dependences}). Upon varying two free parameters in our model, we find that our results agree well with the robust data of open clusters containing FCMDs, such as NGC2547, Pleiades, NGC2516, and Praesepe. Our results reproduce the spins of field stars (with ages greater than about 1 Gyr) within a factor of a few. The agreement can be made more robust with better estimates of inferred ages  {of field stars}. \textcolor{black}{We strengthen the results of works such as \cite{Matt2015} and \cite{Pass2022} by showing that} there is indeed a mass-dependent behaviour of the time-scale over which the spread in spins persists, with more massive FCMDs converging earlier to become slow rotators than less massive FCMDs. This spread lasts up to the {ages of field stars} of about 3 Gyr for FCMDs with $M_\ast=0.35M_\odot$ and longer for less massive FCMDs (Fig.~\ref{fig:1.5&2}). 

\textcolor{black}{We compare and contrast our magnetic braking model results to that of the recently updated calibrated magnetic braking models of \cite{Matt2015}, \cite{Garraffo2018} and to the effect of varying the convective turnover time-scale (Fig.~\ref{fig:diffmod}). With our spin-down model we track the evolution of observationally inferred parameters such as wind mass loss, surface magnetic fields and Alfvén radii, and find a general agreement with observations of the same (Fig.~\ref{fig:obs}).} We shall extend this model to explain the spin evolution of solar-like stars with convective envelopes and radiative cores. Our current model allows us to explore the environment in the vicinity of any exoplanets that may orbit FCMDs.

\section*{Acknowledgements}
\textcolor{black}{The authors thank the referee for their detailed suggestions and comments which greatly enhanced the completeness and readability of the paper.} AS thanks the Gates Cambridge Trust for his scholarship. CAT thanks Churchill College for his fellowship. 
AS thanks Jim Fuller from the California Institute of Technology for suggesting to undertake this work. AS also thanks Patrick Eggenberger from the Université de Genève and Jonathan Irwin from the Institute of Astronomy, University of Cambridge for reading the paper and suggesting edits. AS and CAT thank Dayal Wickramasinghe and Lilia Ferrario for hosting them as visitors at the Australian National University, where the work on dead zones was initiated. AS thanks Francesco Zagaria from the Institute of Astronomy at the University of Cambridge for discussions on the dispersion characteristics of circumstellar disks. AS also thanks Thibault Merle and Lysandra Batail for their hospitality at the Université Libre de Bruxelles, where some of the preliminary results of this work were presented.

\section*{Data availability}
No new data were generated in support of this research. Any numerical codes and related data generated during the work will be made available whenever requested by readers.



\bibliographystyle{mnras}
\bibliography{mnras_template} 

\begin{thebibliography}{}
\makeatletter
\relax
\def\mn@urlcharsother{\let\do\@makeother \do\$\do\&\do\#\do\^\do\_\do\%\do\~}
\def\mn@doi{\begingroup\mn@urlcharsother \@ifnextchar [ {\mn@doi@}
  {\mn@doi@[]}}
\def\mn@doi@[#1]#2{\def\@tempa{#1}\ifx\@tempa\@empty \href
  {http://dx.doi.org/#2} {doi:#2}\else \href {http://dx.doi.org/#2} {#1}\fi
  \endgroup}
\def\mn@eprint#1#2{\mn@eprint@#1:#2::\@nil}
\def\mn@eprint@arXiv#1{\href {http://arxiv.org/abs/#1} {{\tt arXiv:#1}}}
\def\mn@eprint@dblp#1{\href {http://dblp.uni-trier.de/rec/bibtex/#1.xml}
  {dblp:#1}}
\def\mn@eprint@#1:#2:#3:#4\@nil{\def\@tempa {#1}\def\@tempb {#2}\def\@tempc
  {#3}\ifx \@tempc \@empty \let \@tempc \@tempb \let \@tempb \@tempa \fi \ifx
  \@tempb \@empty \def\@tempb {arXiv}\fi \@ifundefined
  {mn@eprint@\@tempb}{\@tempb:\@tempc}{\expandafter \expandafter \csname
  mn@eprint@\@tempb\endcsname \expandafter{\@tempc}}}

\bibitem[\protect\citeauthoryear{Amard, Palacios, Charbonnel, Gallet, Georgy,
  Lagarde  \& Siess}{Amard et~al.}{2019}]{Amard2019}
Amard L.,  Palacios A.,  Charbonnel C.,  Gallet F.,  Georgy C.,  Lagarde N.,
  Siess L.,  2019, \mn@doi [A{\&}A] {10.1051/0004-6361/201935160}, 631, A77

\bibitem[\protect\citeauthoryear{{Barnes}}{{Barnes}}{2003}]{2003ApJ...586..464B}
{Barnes} S.~A.,  2003, \mn@doi [\apj] {10.1086/367639}, \href
  {https://ui.adsabs.harvard.edu/abs/2003ApJ...586..464B} {586, 464}

\bibitem[\protect\citeauthoryear{B{\'{e}}trisey, Eggenberger, Buldgen, Benomar
  \& Bazot}{B{\'{e}}trisey et~al.}{2023}]{Btrisey2023}
B{\'{e}}trisey J.,  Eggenberger P.,  Buldgen G.,  Benomar O.,   Bazot M.,
  2023, \mn@doi [A{\&}A] {10.1051/0004-6361/202245764}, 673, L11

\bibitem[\protect\citeauthoryear{Brown}{Brown}{2014}]{Brown2014}
Brown T.~M.,  2014, \mn@doi [\apj] {10.1088/0004-637x/789/2/101}, 789, 101

\bibitem[\protect\citeauthoryear{Campbell \& Papaloizou}{Campbell \&
  Papaloizou}{1983}]{Campbell1983}
Campbell C.~G.,  Papaloizou J.,  1983, \mn@doi [MNRAS]
  {10.1093/mnras/204.2.433}, 204, 433

\bibitem[\protect\citeauthoryear{Cohen, Glocer, Garraffo, Drake  \& Bell}{Cohen
  et~al.}{2018}]{Cohen2018}
Cohen O.,  Glocer A.,  Garraffo C.,  Drake J.~J.,   Bell J.~M.,  2018, \mn@doi
  [\apj] {10.3847/2041-8213/aab5b5}, 856, L11

\bibitem[\protect\citeauthoryear{{Cowling}}{{Cowling}}{1981}]{1981ARA&A..19..115C}
{Cowling} T.~G.,  1981, \mn@doi [\araa] {10.1146/annurev.aa.19.090181.000555},
  \href {https://ui.adsabs.harvard.edu/abs/1981ARA&A..19..115C} {19, 115}

\bibitem[\protect\citeauthoryear{Douglas, Ag\"{u}eros, Covey, Cargile, Barclay,
  Cody, Howell  \& Kopytova}{Douglas et~al.}{2016}]{Douglas2016}
Douglas S.~T.,  Ag\"{u}eros M.~A.,  Covey K.~R.,  Cargile P.~A.,  Barclay T.,
  Cody A.,  Howell S.~B.,   Kopytova T.,  2016, \mn@doi [\apj]
  {10.3847/0004-637x/822/1/47}, 822, 47

\bibitem[\protect\citeauthoryear{{Durney}}{{Durney}}{1985}]{1985ApJ...297..787D}
{Durney} B.~R.,  1985, \mn@doi [\apj] {10.1086/163575}, \href
  {https://ui.adsabs.harvard.edu/abs/1985ApJ...297..787D} {297, 787}

\bibitem[\protect\citeauthoryear{Duvall, Dziembowski, Goode, Gough, Harvey  \&
  Leibacher}{Duvall et~al.}{1984}]{Duvall1984}
Duvall T.~L.,  Dziembowski W.~A.,  Goode P.~R.,  Gough D.~O.,  Harvey J.~W.,
  Leibacher J.~W.,  1984, \mn@doi [Nature] {10.1038/310022a0}, 310, 22

\bibitem[\protect\citeauthoryear{Eggenberger, Buldgen  \& Salmon}{Eggenberger
  et~al.}{2019}]{Eggenberger2019}
Eggenberger P.,  Buldgen G.,   Salmon S. J. A.~J.,  2019, \mn@doi [A&A]
  {10.1051/0004-6361/201935509}, 626, L1

\bibitem[\protect\citeauthoryear{{Eggleton}}{{Eggleton}}{1973}]{1973MNRAS.163..279E}
{Eggleton} P.~P.,  1973, \mn@doi [\mnras] {10.1093/mnras/163.3.279}, \href
  {https://ui.adsabs.harvard.edu/abs/1973MNRAS.163..279E} {163, 279}

\bibitem[\protect\citeauthoryear{{El-Badry}, {Conroy}, {Fuller}, {Kiman}, {van
  Roestel}, {Rodriguez}  \& {Burdge}}{{El-Badry}
  et~al.}{2022}]{2022MNRAS.517.4916E}
{El-Badry} K.,  {Conroy} C.,  {Fuller} J.,  {Kiman} R.,  {van Roestel} J.,
  {Rodriguez} A.~C.,   {Burdge} K.~B.,  2022, \mn@doi [\mnras]
  {10.1093/mnras/stac2945}, \href
  {https://ui.adsabs.harvard.edu/abs/2022MNRAS.517.4916E} {517, 4916}

\bibitem[\protect\citeauthoryear{{Fedele}, {van den Ancker}, {Henning},
  {Jayawardhana}  \& {Oliveira}}{{Fedele} et~al.}{2010}]{2010A&A...510A..72F}
{Fedele} D.,  {van den Ancker} M.~E.,  {Henning} T.,  {Jayawardhana} R.,
  {Oliveira} J.~M.,  2010, \mn@doi [\aap] {10.1051/0004-6361/200912810}, \href
  {https://ui.adsabs.harvard.edu/abs/2010A&A...510A..72F} {510, A72}

\bibitem[\protect\citeauthoryear{Gallet, Zanni  \& Amard}{Gallet
  et~al.}{2019a}]{Gallet2019}
Gallet F.,  Zanni C.,   Amard L.,  2019a, \mn@doi [A{\&}A]
  {10.1051/0004-6361/201935432}, 632, A6

\bibitem[\protect\citeauthoryear{{Gallet}, {Zanni}  \& {Amard}}{{Gallet}
  et~al.}{2019b}]{2019A&A...632A...6G}
{Gallet} F.,  {Zanni} C.,   {Amard} L.,  2019b, \mn@doi [\aap]
  {10.1051/0004-6361/201935432}, \href
  {https://ui.adsabs.harvard.edu/abs/2019A&A...632A...6G} {632, A6}

\bibitem[\protect\citeauthoryear{G\"{a}nsicke et~al.,}{G\"{a}nsicke
  et~al.}{2009}]{Gnsicke2009}
G\"{a}nsicke B.~T.,  et~al., 2009, \mn@doi [MNRAS]
  {10.1111/j.1365-2966.2009.15126.x}, 397, 2170

\bibitem[\protect\citeauthoryear{Garraffo, Drake  \& Cohen}{Garraffo
  et~al.}{2015}]{Garraffo2015}
Garraffo C.,  Drake J.~J.,   Cohen O.,  2015, \mn@doi [\apj]
  {10.1088/0004-637x/813/1/40}, 813, 40

\bibitem[\protect\citeauthoryear{Garraffo, Drake  \& Cohen}{Garraffo
  et~al.}{2016}]{Garraffo2016}
Garraffo C.,  Drake J.~J.,   Cohen O.,  2016, \mn@doi [A&A]
  {10.1051/0004-6361/201628367}, 595, A110

\bibitem[\protect\citeauthoryear{Garraffo, Drake, Cohen, Alvarado-G{\'{o}}mez
  \& Moschou}{Garraffo et~al.}{2017}]{Garraffo2017}
Garraffo C.,  Drake J.~J.,  Cohen O.,  Alvarado-G{\'{o}}mez J.~D.,   Moschou
  S.~P.,  2017, \mn@doi [\apjl] {10.3847/2041-8213/aa79ed}, 843, L33

\bibitem[\protect\citeauthoryear{Garraffo et~al.,}{Garraffo
  et~al.}{2018a}]{Garraffo2018}
Garraffo C.,  et~al., 2018a, \mn@doi [\apj] {10.3847/1538-4357/aace5d}, 862, 90

\bibitem[\protect\citeauthoryear{{Garraffo}, {Drake}, {Alvarado-Gomez},
  {Moschou}  \& {Cohen}}{{Garraffo} et~al.}{2018b}]{2018ApJ...868...60G}
{Garraffo} C.,  {Drake} J.~J.,  {Alvarado-Gomez} J.~D.,  {Moschou} S.~P.,
  {Cohen} O.,  2018b, \mn@doi [\apj] {10.3847/1538-4357/aae589}, \href
  {https://ui.adsabs.harvard.edu/abs/2018ApJ...868...60G} {868, 60}

\bibitem[\protect\citeauthoryear{Getman, Feigelson  \& Garmire}{Getman
  et~al.}{2023}]{getman2023magnetic}
Getman K.~V.,  Feigelson E.~D.,   Garmire G.~P.,  2023, Magnetic
  Activity-Rotation-Age-Mass Relations in Late Pre-main Sequence Stars
  (\mn@eprint {arXiv} {2305.09013})

\bibitem[\protect\citeauthoryear{Gillon et~al.,}{Gillon
  et~al.}{2017}]{Gillon2017}
Gillon M.,  et~al., 2017, \mn@doi [Nature] {10.1038/nature21360}, 542, 456

\bibitem[\protect\citeauthoryear{{Godoy-Rivera}, {Pinsonneault}  \&
  {Rebull}}{{Godoy-Rivera} et~al.}{2021}]{2021ApJS..257...46G}
{Godoy-Rivera} D.,  {Pinsonneault} M.~H.,   {Rebull} L.~M.,  2021, \mn@doi
  [\apjs] {10.3847/1538-4365/ac2058}, \href
  {https://ui.adsabs.harvard.edu/abs/2021ApJS..257...46G} {257, 46 (G21)}

\bibitem[\protect\citeauthoryear{{Goldreich} \& {Keeley}}{{Goldreich} \&
  {Keeley}}{1977}]{1977ApJ...211..934G}
{Goldreich} P.,  {Keeley} D.~A.,  1977, \mn@doi [\apj] {10.1086/155005}, \href
  {https://ui.adsabs.harvard.edu/abs/1977ApJ...211..934G} {211, 934}

\bibitem[\protect\citeauthoryear{Gossage, Dotter, Garraffo, Drake, Douglas  \&
  Conroy}{Gossage et~al.}{2021}]{Gossage2021}
Gossage S.,  Dotter A.,  Garraffo C.,  Drake J.~J.,  Douglas S.,   Conroy C.,
  2021, \mn@doi [ApJ] {10.3847/1538-4357/abebdf}, 912, 65

\bibitem[\protect\citeauthoryear{{Hillenbrand}}{{Hillenbrand}}{1997}]{1997AJ....113.1733H}
{Hillenbrand} L.~A.,  1997, \mn@doi [\aj] {10.1086/118389}, \href
  {https://ui.adsabs.harvard.edu/abs/1997AJ....113.1733H} {113, 1733}

\bibitem[\protect\citeauthoryear{{Ireland}, {Zanni}, {Matt}  \&
  {Pantolmos}}{{Ireland} et~al.}{2021}]{2021ApJ...906....4I}
{Ireland} L.~G.,  {Zanni} C.,  {Matt} S.~P.,   {Pantolmos} G.,  2021, \mn@doi
  [\apj] {10.3847/1538-4357/abc828}, \href
  {https://ui.adsabs.harvard.edu/abs/2021ApJ...906....4I} {906, 4}

\bibitem[\protect\citeauthoryear{Jao, Henry, White, Nisak, Hubbard-James  \&
  Paredes}{Jao et~al.}{2023}]{jao2023mind}
Jao W.-C.,  Henry T.~J.,  White R.~J.,  Nisak A.~H.,  Hubbard-James H.-S.,
  Paredes L.~A.,  2023, Mind the Gap I: H$\alpha$ Activity of M Dwarfs Near the
  Partially/Fully Convective Boundary and a New H$\alpha$ Emission Deficiency
  Zone on the Main Sequence (\mn@eprint {arXiv} {2304.14452})

\bibitem[\protect\citeauthoryear{{Jerabkova}, {Beccari}, {Boffin},
  {Petr-Gotzens}, {Manara}, {Prada Moroni}, {Tognelli}  \&
  {Degl'Innocenti}}{{Jerabkova} et~al.}{2019}]{2019A&A...627A..57J}
{Jerabkova} T.,  {Beccari} G.,  {Boffin} H. M.~J.,  {Petr-Gotzens} M.~G.,
  {Manara} C.~F.,  {Prada Moroni} P.~G.,  {Tognelli} E.,   {Degl'Innocenti} S.,
   2019, \mn@doi [\aap] {10.1051/0004-6361/201935016}, \href
  {https://ui.adsabs.harvard.edu/abs/2019A&A...627A..57J} {627, A57}

\bibitem[\protect\citeauthoryear{{Kawaler}}{{Kawaler}}{1988}]{1988ApJ...333..236K}
{Kawaler} S.~D.,  1988, \mn@doi [\apj] {10.1086/166740}, \href
  {https://ui.adsabs.harvard.edu/abs/1988ApJ...333..236K} {333, 236}

\bibitem[\protect\citeauthoryear{{Kivelson} \& {Ridley}}{{Kivelson} \&
  {Ridley}}{2008}]{2008JGRA..113.5214K}
{Kivelson} M.~G.,  {Ridley} A.~J.,  2008, \mn@doi [Journal of Geophysical
  Research (Space Physics)] {10.1029/2007JA012302}, \href
  {https://ui.adsabs.harvard.edu/abs/2008JGRA..113.5214K} {113, A05214}

\bibitem[\protect\citeauthoryear{Knigge}{Knigge}{2006}]{Knigge2006}
Knigge C.,  2006, \mn@doi [MNRAS] {10.1111/j.1365-2966.2006.11096.x}, 373, 484

\bibitem[\protect\citeauthoryear{Knigge, Baraffe  \& Patterson}{Knigge
  et~al.}{2011}]{Knigge2011}
Knigge C.,  Baraffe I.,   Patterson J.,  2011, \mn@doi [ApJS]
  {10.1088/0067-0049/194/2/28}, 194, 28

\bibitem[\protect\citeauthoryear{{Kochukhov}}{{Kochukhov}}{2021}]{2021A&ARv..29....1K}
{Kochukhov} O.,  2021, \mn@doi [\aapr] {10.1007/s00159-020-00130-3}, \href
  {https://ui.adsabs.harvard.edu/abs/2021A&ARv..29....1K} {29, 1}

\bibitem[\protect\citeauthoryear{{Kolb}}{{Kolb}}{1993}]{1993A&A...271..149K}
{Kolb} U.,  1993, \aap, \href
  {https://ui.adsabs.harvard.edu/abs/1993A&A...271..149K} {271, 149}

\bibitem[\protect\citeauthoryear{{Kolb} \& {Baraffe}}{{Kolb} \&
  {Baraffe}}{1999}]{1999MNRAS.309.1034K}
{Kolb} U.,  {Baraffe} I.,  1999, \mn@doi [\mnras]
  {10.1046/j.1365-8711.1999.02926.x}, \href
  {https://ui.adsabs.harvard.edu/abs/1999MNRAS.309.1034K} {309, 1034}

\bibitem[\protect\citeauthoryear{Kounkel, Stassun, Hillenbrand,
  Hern{\'{a}}ndez, Serna  \& Curtis}{Kounkel et~al.}{2023}]{Kounkel2023}
Kounkel M.,  Stassun K.~G.,  Hillenbrand L.~A.,  Hern{\'{a}}ndez J.,  Serna J.,
    Curtis J.~L.,  2023, \mn@doi [\aj] {10.3847/1538-3881/acc2bd}, 165, 182

\bibitem[\protect\citeauthoryear{{Li}, {Wu}  \& {Wickramasinghe}}{{Li}
  et~al.}{1994}]{1994MNRAS.268...61L}
{Li} J.~K.,  {Wu} K.~W.,   {Wickramasinghe} D.~T.,  1994, \mn@doi [\mnras]
  {10.1093/mnras/268.1.61}, \href
  {https://ui.adsabs.harvard.edu/abs/1994MNRAS.268...61L} {268, 61}

\bibitem[\protect\citeauthoryear{{Marcy} \& {Chen}}{{Marcy} \&
  {Chen}}{1992}]{1992ApJ...390..550M}
{Marcy} G.~W.,  {Chen} G.~H.,  1992, \mn@doi [\apj] {10.1086/171305}, \href
  {https://ui.adsabs.harvard.edu/abs/1992ApJ...390..550M} {390, 550}

\bibitem[\protect\citeauthoryear{Matt, Brun, Baraffe, Bouvier  \&
  Chabrier}{Matt et~al.}{2015}]{Matt2015}
Matt S.~P.,  Brun A.~S.,  Baraffe I.,  Bouvier J.,   Chabrier G.,  2015,
  \mn@doi [\apj] {10.1088/2041-8205/799/2/l23}, 799, L23

\bibitem[\protect\citeauthoryear{{Meibom}, {Mathieu}  \& {Stassun}}{{Meibom}
  et~al.}{2009}]{2009ApJ...695..679M}
{Meibom} S.,  {Mathieu} R.~D.,   {Stassun} K.~G.,  2009, \mn@doi [\apj]
  {10.1088/0004-637X/695/1/679}, \href
  {https://ui.adsabs.harvard.edu/abs/2009ApJ...695..679M} {695, 679}

\bibitem[\protect\citeauthoryear{{Meibom} et~al.,}{{Meibom}
  et~al.}{2011}]{2011ApJ...733L...9M}
{Meibom} S.,  et~al., 2011, \mn@doi [\apjl] {10.1088/2041-8205/733/1/L9}, \href
  {https://ui.adsabs.harvard.edu/abs/2011ApJ...733L...9M} {733, L9}

\bibitem[\protect\citeauthoryear{{Mestel} \& {Spruit}}{{Mestel} \&
  {Spruit}}{1987}]{1987MNRAS.226...57M}
{Mestel} L.,  {Spruit} H.~C.,  1987, \mn@doi [\mnras] {10.1093/mnras/226.1.57},
  \href {https://ui.adsabs.harvard.edu/abs/1987MNRAS.226...57M} {226, 57 (MS)}

\bibitem[\protect\citeauthoryear{{Metcalfe} et~al.,}{{Metcalfe}
  et~al.}{2022}]{2022ApJ...933L..17M}
{Metcalfe} T.~S.,  et~al., 2022, \mn@doi [\apjl] {10.3847/2041-8213/ac794d},
  \href {https://ui.adsabs.harvard.edu/abs/2022ApJ...933L..17M} {933, L17}

\bibitem[\protect\citeauthoryear{{Metcalfe} et~al.,}{{Metcalfe}
  et~al.}{2023}]{2023ApJ...948L...6M}
{Metcalfe} T.~S.,  et~al., 2023, \mn@doi [\apjl] {10.3847/2041-8213/acce38},
  \href {https://ui.adsabs.harvard.edu/abs/2023ApJ...948L...6M} {948, L6}

\bibitem[\protect\citeauthoryear{{Morin}, {Dormy}, {Schrinner}  \&
  {Donati}}{{Morin} et~al.}{2011}]{2011MNRAS.418L.133M}
{Morin} J.,  {Dormy} E.,  {Schrinner} M.,   {Donati} J.~F.,  2011, \mn@doi
  [\mnras] {10.1111/j.1745-3933.2011.01159.x}, \href
  {https://ui.adsabs.harvard.edu/abs/2011MNRAS.418L.133M} {418, L133}

\bibitem[\protect\citeauthoryear{Newton, Irwin, Charbonneau, Berta-Thompson,
  Dittmann  \& West}{Newton et~al.}{2016}]{Newton2016}
Newton E.~R.,  Irwin J.,  Charbonneau D.,  Berta-Thompson Z.~K.,  Dittmann
  J.~A.,   West A.~A.,  2016, \mn@doi [\apj] {10.3847/0004-637x/821/2/93}, 821,
  93

\bibitem[\protect\citeauthoryear{{Paczy{\'n}ski}}{{Paczy{\'n}ski}}{1967}]{1967AcA....17..287P}
{Paczy{\'n}ski} B.,  1967, \actaa, \href
  {https://ui.adsabs.harvard.edu/abs/1967AcA....17..287P} {17, 287}

\bibitem[\protect\citeauthoryear{Pala et~al.,}{Pala et~al.}{2021}]{Pala2021}
Pala A.~F.,  et~al., 2021, \mn@doi [\mnras] {10.1093/mnras/stab3449}, 510, 6110

\bibitem[\protect\citeauthoryear{{Pantolmos}, {Zanni}  \&
  {Bouvier}}{{Pantolmos} et~al.}{2020}]{2020A&A...643A.129P}
{Pantolmos} G.,  {Zanni} C.,   {Bouvier} J.,  2020, \mn@doi [\aap]
  {10.1051/0004-6361/202038569}, \href
  {https://ui.adsabs.harvard.edu/abs/2020A&A...643A.129P} {643, A129}

\bibitem[\protect\citeauthoryear{{Parker}}{{Parker}}{1955}]{1955ApJ...122..293P}
{Parker} E.~N.,  1955, \mn@doi [\apj] {10.1086/146087}, \href
  {https://ui.adsabs.harvard.edu/abs/1955ApJ...122..293P} {122, 293}

\bibitem[\protect\citeauthoryear{Pass, Charbonneau, Irwin  \& Winters}{Pass
  et~al.}{2022}]{Pass2022}
Pass E.~K.,  Charbonneau D.,  Irwin J.~M.,   Winters J.~G.,  2022, \mn@doi
  [\apj] {10.3847/1538-4357/ac7da8}, 936, 109

\bibitem[\protect\citeauthoryear{{Pols}, {Tout}, {Eggleton}  \& {Han}}{{Pols}
  et~al.}{1995}]{1995MNRAS.274..964P}
{Pols} O.~R.,  {Tout} C.~A.,  {Eggleton} P.~P.,   {Han} Z.,  1995, \mn@doi
  [\mnras] {10.1093/mnras/274.3.964}, \href
  {https://ui.adsabs.harvard.edu/abs/1995MNRAS.274..964P} {274, 964}

\bibitem[\protect\citeauthoryear{Popinchalk, Faherty, Kiman, Gagn{\'{e}},
  Curtis, Angus, Cruz  \& Rice}{Popinchalk et~al.}{2021}]{Popinchalk2021}
Popinchalk M.,  Faherty J.~K.,  Kiman R.,  Gagn{\'{e}} J.,  Curtis J.~L.,
  Angus R.,  Cruz K.~L.,   Rice E.~L.,  2021, \mn@doi [\apj]
  {10.3847/1538-4357/ac0444}, 916, 77

\bibitem[\protect\citeauthoryear{{Pylyser} \& {Savonije}}{{Pylyser} \&
  {Savonije}}{1988}]{1988A&A...191...57P}
{Pylyser} E.,  {Savonije} G.~J.,  1988, \aap, \href
  {https://ui.adsabs.harvard.edu/abs/1988A&A...191...57P} {191, 57}

\bibitem[\protect\citeauthoryear{Rebull, Wolff  \& Strom}{Rebull
  et~al.}{2004}]{Rebull2004}
Rebull L.~M.,  Wolff S.~C.,   Strom S.~E.,  2004, \mn@doi [\aj]
  {10.1086/380931}, 127, 1029

\bibitem[\protect\citeauthoryear{{Rebull}, {Stauffer}, {Cody}, {Hillenbrand},
  {David}  \& {Pinsonneault}}{{Rebull} et~al.}{2018}]{2018AJ....155..196R}
{Rebull} L.~M.,  {Stauffer} J.~R.,  {Cody} A.~M.,  {Hillenbrand} L.~A.,
  {David} T.~J.,   {Pinsonneault} M.,  2018, \mn@doi [\aj]
  {10.3847/1538-3881/aab605}, \href
  {https://ui.adsabs.harvard.edu/abs/2018AJ....155..196R} {155, 196}

\bibitem[\protect\citeauthoryear{{Reg{\H{o}}s} \& {Tout}}{{Reg{\H{o}}s} \&
  {Tout}}{1995}]{1995MNRAS.273..146R}
{Reg{\H{o}}s} E.,  {Tout} C.~A.,  1995, \mn@doi [\mnras]
  {10.1093/mnras/273.1.146}, \href
  {https://ui.adsabs.harvard.edu/abs/1995MNRAS.273..146R} {273, 146}

\bibitem[\protect\citeauthoryear{{Reiners} \& {Mohanty}}{{Reiners} \&
  {Mohanty}}{2012}]{2012ApJ...746...43R}
{Reiners} A.,  {Mohanty} S.,  2012, \mn@doi [\apj]
  {10.1088/0004-637X/746/1/43}, \href
  {https://ui.adsabs.harvard.edu/abs/2012ApJ...746...43R} {746, 43}

\bibitem[\protect\citeauthoryear{{Reiners}, {Basri}  \& {Browning}}{{Reiners}
  et~al.}{2009}]{2009ApJ...692..538R}
{Reiners} A.,  {Basri} G.,   {Browning} M.,  2009, \mn@doi [\apj]
  {10.1088/0004-637X/692/1/538}, \href
  {https://ui.adsabs.harvard.edu/abs/2009ApJ...692..538R} {692, 538}

\bibitem[\protect\citeauthoryear{{R{\'e}ville}, {Brun}, {Matt}, {Strugarek}  \&
  {Pinto}}{{R{\'e}ville} et~al.}{2015}]{2015ApJ...798..116R}
{R{\'e}ville} V.,  {Brun} A.~S.,  {Matt} S.~P.,  {Strugarek} A.,   {Pinto}
  R.~F.,  2015, \mn@doi [\apj] {10.1088/0004-637X/798/2/116}, \href
  {https://ui.adsabs.harvard.edu/abs/2015ApJ...798..116R} {798, 116}

\bibitem[\protect\citeauthoryear{{Roquette}, {Matt}, {Winter}, {Amard}  \&
  {Stasevic}}{{Roquette} et~al.}{2021}]{2021MNRAS.508.3710R}
{Roquette} J.,  {Matt} S.~P.,  {Winter} A.~J.,  {Amard} L.,   {Stasevic} S.,
  2021, \mn@doi [\mnras] {10.1093/mnras/stab2772}, \href
  {https://ui.adsabs.harvard.edu/abs/2021MNRAS.508.3710R} {508, 3710}

\bibitem[\protect\citeauthoryear{{Sarkar} \& {Tout}}{{Sarkar} \&
  {Tout}}{2022}]{2022MNRAS.513.4169S}
{Sarkar} A.,  {Tout} C.~A.,  2022, \mn@doi [\mnras] {10.1093/mnras/stac1187},
  \href {https://ui.adsabs.harvard.edu/abs/2022MNRAS.513.4169S} {513, 4169}

\bibitem[\protect\citeauthoryear{{Sarkar}, {Ge}  \& {Tout}}{{Sarkar}
  et~al.}{2023a}]{2023MNRAS.519.2567S}
{Sarkar} A.,  {Ge} H.,   {Tout} C.~A.,  2023a, \mn@doi [\mnras]
  {10.1093/mnras/stac3688}, \href
  {https://ui.adsabs.harvard.edu/abs/2023MNRAS.519.2567S} {519, 2567}

\bibitem[\protect\citeauthoryear{Sarkar, Ge  \& Tout}{Sarkar
  et~al.}{2023b}]{Sarkar2023}
Sarkar A.,  Ge H.,   Tout C.~A.,  2023b, \mn@doi [\mnras]
  {10.1093/mnras/stad354}, 520, 3187

\bibitem[\protect\citeauthoryear{See et~al.,}{See et~al.}{2016}]{See2016}
See V.,  et~al., 2016, \mn@doi [\mnras] {10.1093/mnras/stw3094}, 466, 1542

\bibitem[\protect\citeauthoryear{{See} et~al.,}{{See}
  et~al.}{2019}]{2019ApJ...886..120S}
{See} V.,  et~al., 2019, \mn@doi [\apj] {10.3847/1538-4357/ab46b2}, \href
  {https://ui.adsabs.harvard.edu/abs/2019ApJ...886..120S} {886, 120}

\bibitem[\protect\citeauthoryear{Sills, Pinsonneault  \& Terndrup}{Sills
  et~al.}{2000}]{Sills2000}
Sills A.,  Pinsonneault M.~H.,   Terndrup D.~M.,  2000, \mn@doi [\apj]
  {10.1086/308739}, 534, 335

\bibitem[\protect\citeauthoryear{{Skumanich}}{{Skumanich}}{1972}]{1972ApJ...171..565S}
{Skumanich} A.,  1972, \mn@doi [\apj] {10.1086/151310}, \href
  {https://ui.adsabs.harvard.edu/abs/1972ApJ...171..565S} {171, 565}

\bibitem[\protect\citeauthoryear{{Spruit}}{{Spruit}}{2002}]{2002A&A...381..923S}
{Spruit} H.~C.,  2002, \mn@doi [\aap] {10.1051/0004-6361:20011465}, \href
  {https://ui.adsabs.harvard.edu/abs/2002A&A...381..923S} {381, 923}

\bibitem[\protect\citeauthoryear{{Stahler}}{{Stahler}}{1983}]{1983ApJ...274..822S}
{Stahler} S.~W.,  1983, \mn@doi [\apj] {10.1086/161495}, \href
  {https://ui.adsabs.harvard.edu/abs/1983ApJ...274..822S} {274, 822}

\bibitem[\protect\citeauthoryear{{Taam} \& {Spruit}}{{Taam} \&
  {Spruit}}{1989}]{1989ApJ...345..972T}
{Taam} R.~E.,  {Spruit} H.~C.,  1989, \mn@doi [\apj] {10.1086/167966}, \href
  {https://ui.adsabs.harvard.edu/abs/1989ApJ...345..972T} {345, 972}

\bibitem[\protect\citeauthoryear{{Tout} \& {Pringle}}{{Tout} \&
  {Pringle}}{1992}]{1992MNRAS.256..269T}
{Tout} C.~A.,  {Pringle} J.~E.,  1992, \mn@doi [\mnras]
  {10.1093/mnras/256.2.269}, \href
  {https://ui.adsabs.harvard.edu/abs/1992MNRAS.256..269T} {256, 269 (TP)}

\bibitem[\protect\citeauthoryear{{Tutukov}, {Fedorova}, {Ergma}  \&
  {Yungelson}}{{Tutukov} et~al.}{1985}]{1985SvAL...11...52T}
{Tutukov} A.~V.,  {Fedorova} A.~V.,  {Ergma} E.~V.,   {Yungelson} L.~R.,  1985,
  Soviet Astronomy Letters, \href
  {https://ui.adsabs.harvard.edu/abs/1985SvAL...11...52T} {11, 52}

\bibitem[\protect\citeauthoryear{{Verbunt} \& {Zwaan}}{{Verbunt} \&
  {Zwaan}}{1981}]{1981A&A...100L...7V}
{Verbunt} F.,  {Zwaan} C.,  1981, \aap, \href
  {https://ui.adsabs.harvard.edu/abs/1981A&A...100L...7V} {100, L7}

\bibitem[\protect\citeauthoryear{{Villarreal D'Angelo}, {Jardine}  \&
  {See}}{{Villarreal D'Angelo} et~al.}{2018}]{2018MNRAS.475L..25V}
{Villarreal D'Angelo} C.,  {Jardine} M.,   {See} V.,  2018, \mn@doi [\mnras]
  {10.1093/mnrasl/slx206}, \href
  {https://ui.adsabs.harvard.edu/abs/2018MNRAS.475L..25V} {475, L25}

\bibitem[\protect\citeauthoryear{{Warner}}{{Warner}}{2003}]{2003cvs..book.....W}
{Warner} B.,  2003, {Cataclysmic Variable Stars}.
Cambridge University Press, \mn@doi{10.1017/CBO9780511586491}

\bibitem[\protect\citeauthoryear{{Weber} \& {Davis}}{{Weber} \&
  {Davis}}{1967}]{1967ApJ...148..217W}
{Weber} E.~J.,  {Davis} Leverett J.,  1967, \mn@doi [\apj] {10.1086/149138},
  \href {https://ui.adsabs.harvard.edu/abs/1967ApJ...148..217W} {148, 217}

\bibitem[\protect\citeauthoryear{Wijnen, Zorotovic  \& Schreiber}{Wijnen
  et~al.}{2015}]{Wijnen2015}
Wijnen T. P.~G.,  Zorotovic M.,   Schreiber M.~R.,  2015, \mn@doi [A{\&}A]
  {10.1051/0004-6361/201323018}, 577, A143

\bibitem[\protect\citeauthoryear{{Wright}, {Drake}, {Mamajek}  \&
  {Henry}}{{Wright} et~al.}{2011}]{2011ApJ...743...48W}
{Wright} N.~J.,  {Drake} J.~J.,  {Mamajek} E.~E.,   {Henry} G.~W.,  2011,
  \mn@doi [\apj] {10.1088/0004-637X/743/1/48}, \href
  {https://ui.adsabs.harvard.edu/abs/2011ApJ...743...48W} {743, 48}

\bibitem[\protect\citeauthoryear{{Wright}, {Newton}, {Williams}, {Drake}  \&
  {Yadav}}{{Wright} et~al.}{2018}]{2018MNRAS.479.2351W}
{Wright} N.~J.,  {Newton} E.~R.,  {Williams} P. K.~G.,  {Drake} J.~J.,
  {Yadav} R.~K.,  2018, \mn@doi [\mnras] {10.1093/mnras/sty1670}, \href
  {https://ui.adsabs.harvard.edu/abs/2018MNRAS.479.2351W} {479, 2351}

\bibitem[\protect\citeauthoryear{Zangrilli, Tout  \& Bianchini}{Zangrilli
  et~al.}{1997}]{Zangrilli1997}
Zangrilli L.,  Tout C.~A.,   Bianchini A.,  1997, \mn@doi [\mnras]
  {10.1093/mnras/289.1.59}, 289, 59

\bibitem[\protect\citeauthoryear{{Zhang}, {Wickramasinghe}  \&
  {Ferrario}}{{Zhang} et~al.}{2009}]{2009MNRAS.397.2208Z}
{Zhang} C.~M.,  {Wickramasinghe} D.~T.,   {Ferrario} L.,  2009, \mn@doi
  [\mnras] {10.1111/j.1365-2966.2009.15154.x}, \href
  {https://ui.adsabs.harvard.edu/abs/2009MNRAS.397.2208Z} {397, 2208}

\bibitem[\protect\citeauthoryear{van Saders, Pinsonneault  \& Barbieri}{van
  Saders et~al.}{2019}]{vanSaders2019}
van Saders J.~L.,  Pinsonneault M.~H.,   Barbieri M.,  2019, \mn@doi [\apj]
  {10.3847/1538-4357/aafafe}, 872, 128

\makeatother
\end{thebibliography}



\appendix

\section{Effect of metallicity}
\begin{figure}
\includegraphics[width=0.45\textwidth]{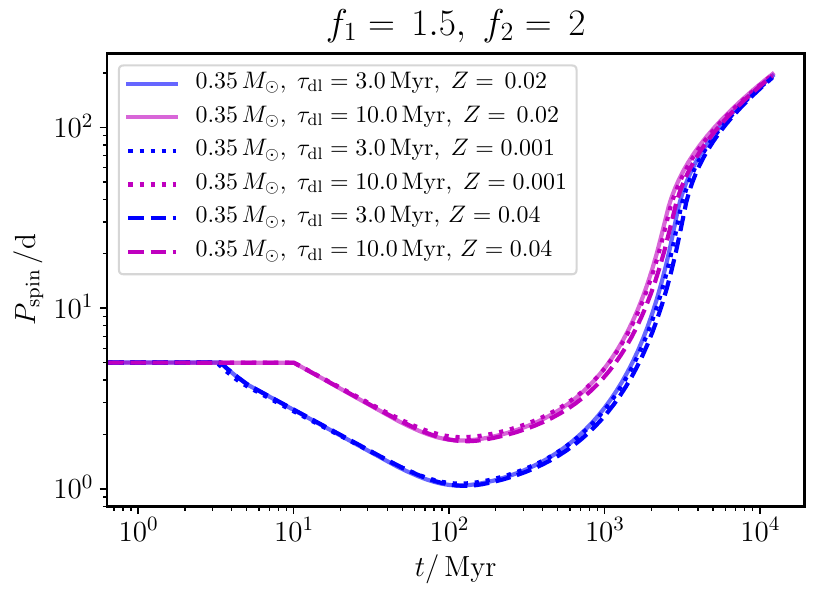}
\caption{The dependence of our model trajectories on metallicity $Z$ for a set of trajectories with $M_\ast=0.35\,M_\odot$, initial $P_\mathrm{spin}=\,5\,\mathrm{d}$, $\tau_\mathrm{dl}/\,\mathrm{Myr} \in \{3,\,10\}$, and $Z\in\{0.001,\,0.02,\,0.04\}$.}
\label{fig:Z}
\end{figure}

We also assess the dependence of our model trajectories on the stellar metallicity $Z$. We plot a set of trajectories with $M_\ast=0.35\,M_\odot$, initial $P_\mathrm{spin}=\,5\,\mathrm{d}$, $\tau_\mathrm{dl}/\,\mathrm{Myr} \in \{3,\,10\}$ and $Z\in\{0.001,\,0.02,\,0.04\}$. Our results are shown in Fig.~\ref{fig:Z}. We see that the spin evolution of our trajectories has a negligible dependence on the metallicity.

\bsp	
\label{lastpage}
\end{document}